\documentclass[10pt]{article}

\usepackage[letterpaper,margin=0.72in]{geometry}
\usepackage[T1]{fontenc}
\usepackage[utf8]{inputenc}
\usepackage{lmodern}
\usepackage{microtype}
\usepackage{amsmath,amssymb}
\usepackage{graphicx}
\usepackage{array}
\usepackage{booktabs}
\usepackage{longtable}
\usepackage{pdflscape}
\usepackage{caption}
\usepackage{float}
\usepackage{placeins}
\usepackage{xurl}
\usepackage[colorlinks=true,linkcolor=blue,citecolor=blue,urlcolor=blue]{hyperref}

\newcommand{\lpact}{L-PACT}
\newcommand{\code}[1]{\nolinkurl{#1}}
\newcommand{\yes}{yes}
\newcommand{\no}{no}
\newcolumntype{L}[1]{>{\raggedright\arraybackslash}p{#1}}
\newcolumntype{C}[1]{>{\centering\arraybackslash}p{#1}}
\title{Do Language Models Align with Brains? Prediction Scores Are Not Enough}
\author{Xiao Jia\\School of Artificial Intelligence, The Chinese University of Hong Kong, Shenzhen, China\\\texttt{xiaojia@link.cuhk.edu.cn}}
\date{\today}

\begin{document}
\maketitle

\begin{abstract}
Brain-language model comparisons often interpret neural prediction scores as evidence that model representations capture brain-relevant language computation. We asked whether language models align with brains, and whether prediction scores are enough to support that claim, using L-PACT, a source-audited framework that evaluates predictive, relational, mechanism-stripping, and reliability-bounded evidence. Across primary naturalistic language neural datasets and derived language-model representations, L-PACT compared real model features with nuisance baselines and severe controls, tested whether model-to-brain profiles reproduced brain-to-brain patterns, recomputed held-out scores after mechanism stripping, and normalized evidence against brain-brain ceilings. The locked analysis set contains 414 predictive-control rows, 2304 relational profile rows, 4320 mechanism-stripping rows, 420 brain-brain ceiling rows, and 146 integrated decision rows. Assay-sensitivity checks showed that brain-brain reliability, brain-as-model run-to-run relational profiles, independent low-level neural and WAV-derived acoustic-envelope gates, and a deterministic implanted-signal simulation can produce positive evidence when expected. Nevertheless, no real model row passed the predictive, relational, mechanism-stripping, or operational Turing-bounded reliability gates; all 146 integrated rows were control-explained. Less stringent single-criterion rules would have counted raw positive predictive, relational, stripping-delta, and ceiling-normalized effects, but L-PACT downgraded them because controls explained the apparent evidence. In the analyzed derived artifact set, the tested language-model representations do not satisfy L-PACT alignment gates; apparent positives are converted into an auditable control-explained taxonomy rather than treated as structural alignment.
\end{abstract}

\noindent\textbf{Significance statement.} A central question in NeuroAI is whether language models align with brains, rather than merely producing useful neural prediction scores. L-PACT separates prediction-score evidence from relational, mechanism-stripping, and reliability-bounded alignment claims. Brain-derived, low-level neural, acoustic-envelope, and implanted-signal positive controls show that the assay can detect structure when it is present. In the current source-audited analysis, apparent model-brain correspondences are nevertheless control-explained. This distinction matters because prediction scores alone can make computational models seem more biologically informative than the evidence supports.


\section*{Introduction}

Large language models have become central objects in computational neuroscience because their internal representations provide high-dimensional predictors for neural measurements collected during language comprehension. Studies now routinely compare model features with fMRI, ECoG, MEG, eye tracking, and behavioral data in order to ask whether model representations capture brain-relevant structure \cite{huth2016,jain2018,schrimpf2021,caucheteux2022,goldstein2022,toneva2019,tuckute2024,pereira2018,wehbe2014,lerner2011,fedorenko2011,blank2016,brennan2016,ding2016,brodbeck2018,broderick2018,brainscore,yamins2014}. Such comparisons are valuable, but they also create an evidential ambiguity. A model feature can predict a neural response because it carries stimulus timing, lexical frequency, contextual regularity, dimensional structure, or autocorrelation. Predictive utility is therefore not identical to structural or mechanistic correspondence.

The model side of this comparison has also changed. Static lexical representations gave way to contextual transformer models whose features are layer-indexed and scale-dependent \cite{mikolov2013,pennington2014,vaswani2017,devlin2019,radford2019,brown2020,kaplan2020,hoffmann2022,biderman2023,qwen25,qwen3}. This capacity makes neural prediction easier to obtain but harder to interpret, increasing the need to separate useful statistical features from evidence of relational organization or mechanism-specific dependence.

The terminology in this paper is therefore deliberately asymmetric. We use \emph{correspondence} for any reproducible statistical association between model-derived quantities and neural measurements, including associations that may be useful for prediction or experimental annotation. We reserve \emph{alignment} for a stronger, gate-qualified claim that the association is control-robust, relationally organized, mechanism-specific, and interpretable relative to brain-brain reliability. This distinction prevents a positive local encoding score from carrying the evidential burden of a network-level or mechanism-level claim.

The present work addresses this ambiguity by distinguishing four evidence levels. Predictive adequacy asks whether model-derived features improve held-out neural prediction relative to nuisance baselines and severe controls. Relational adequacy asks whether model-to-brain profiles reproduce brain-to-brain alignment patterns, rather than only improving a local score. Counterfactual mechanism-stripping adequacy asks whether removing a candidate mechanism from the model representation produces selective held-out deficits for matching neural targets within the implemented predictor. Reliability-bounded adequacy asks whether model evidence approaches the variability observed among brain measurements themselves. These levels are related, but none is logically interchangeable with another.

This distinction is particularly important for brain-language model evaluation. Naturalistic language stimuli contain strong temporal and lexical structure, and modern model representations are high-dimensional enough to support flexible readouts \cite{naselaris2011,hastie2009,stone1974}. Without severe controls, a positive neural prediction score may be overinterpreted as evidence that a model implements the organization of the language system. Conversely, a model may be useful as a predictor while not satisfying stricter tests of relational structure, mechanism-specific necessity, or brain-brain reliability. A rigorous evaluation framework should preserve useful predictive evidence while preventing it from being converted into stronger claims than the tests justify.

We therefore introduce L-PACT, a Language Predictive, Alignment-pattern, Causal, and Turing-bounded Test, to test whether prediction scores are sufficient for alignment claims by separating them from increasingly stronger evidence levels. The acronym retains ``Causal'' as part of the established L-PACT name; in this manuscript, the causal component is operationalized as counterfactual mechanism stripping within the model-based predictor, not biological intervention. The Turing-bounded component is operationalized as reliability-bounded interpretation relative to available brain-brain estimates, not as a claim that the NeuroAI Turing Test is a settled field standard. The relational component follows the logic that a model comparison should test cross-unit organization rather than only local scores \cite{kriegeskorte2008,kornblith2019,apa}, while the reliability-bounded component follows the broader NeuroAI argument that model evidence should be interpreted relative to brain-to-brain variability \cite{neuroaituring}. This study is a source-audited reanalysis of derived neural and model artifacts. It tests evidence standards rather than introducing a new neural acquisition dataset or model comparison benchmark. The study evaluates derived artifacts from Brain Treebank, Podcast ECoG, MEG-MASC, and additional secondary or diagnostic sources. NaturalStories is explicitly excluded from primary neural evidence because the available artifact chain does not satisfy the configured neural-side evidence requirements.

Interpreting a negative-control outcome requires evidence that the assay can detect known structure. We therefore added positive controls that do not change the real model decisions: brain-brain reliability controls ask whether the derived neural artifacts contain stable brain-derived structure; brain-as-model relational controls ask whether held-out brain profiles can pass a L-PACT-like profile gate; low-level neural sanity gates ask whether word-onset or word-rate targets can be detected independently of language-model features; standalone acoustic-envelope gates ask whether local WAV-derived acoustic structure predicts neural time series beyond temporal controls; and a deterministic implanted-signal simulation asks whether the gate architecture can pass when predictive, relational, mechanism-stripping, and ceiling-bounded signal is known to be present. These controls separate failures of the tested model representations from failures of the assay itself.

The locked analysis set contains 414 predictive-control rows, 2304 relational profile rows, 4320 mechanism-stripping rows, 420 brain-brain ceiling rows, and 146 integrated decision rows. No real model row passes the predictive, relational, mechanism-stripping, or operational Turing-bounded reliability gates. All 146 integrated decision rows are control-explained under L-PACT gates, and no row qualified as an integrated L-PACT candidate. The result is therefore not a general verdict on whether current language models can ever align with brains. It is more conservative and more specific: less stringent analyses can yield prediction-looking positives, but prediction scores are not enough for alignment because the analyzed derived artifact set does not support positive structural or mechanism-specific evidence and prediction-like evidence is control-explained under severe controls.

The contribution is not a new language model and not a claim that all possible model-brain comparisons are negative. It is an evidence architecture and a source-audited empirical evaluation: L-PACT specifies what stronger claims would require, checks which parts of the evidence chain are available, and records when severe controls explain the apparent effect. This makes the control-explained result actionable for future work, because a future positive claim can be localized to the gate it newly satisfies rather than asserted from an aggregate score.

\section*{Results}

\begin{figure}[htbp]
\centering
\includegraphics[width=\textwidth]{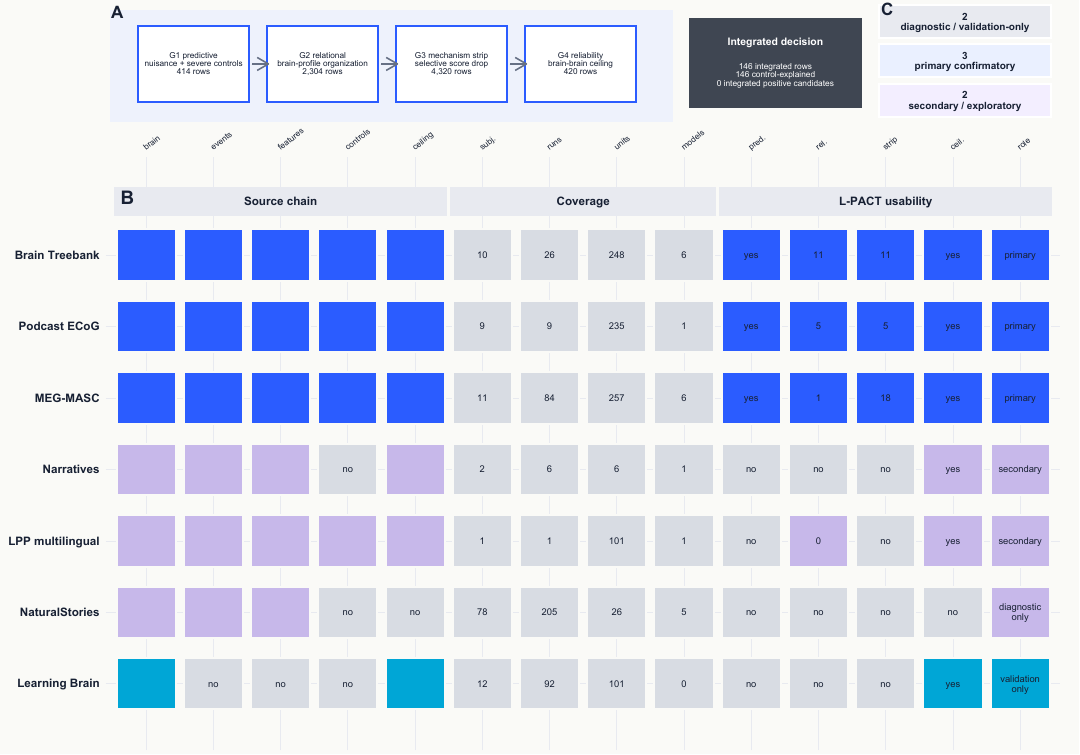}
\caption{L-PACT framework and source-audited dataset eligibility. (A) The evidence hierarchy separates predictive adequacy, relational adequacy, counterfactual mechanism-stripping adequacy, and reliability-bounded adequacy before assigning an integrated decision. The locked analysis set contains 414 predictive-control rows, 2304 relational profile rows, 4320 mechanism-stripping rows, 420 brain-brain ceiling rows, and 146 integrated decision rows; all integrated rows are control-explained. (B) Source-audited dataset eligibility distinguishes primary confirmatory datasets from secondary, diagnostic, or validation-only sources. NaturalStories entries are diagnostic/stimulus-only and excluded from primary L-PACT inference. (C) Dataset-role summary.}
\label{fig:lpact_framework}
\end{figure}

\textbf{L-PACT separates predictive, relational, mechanism-stripping, and reliability-bounded evidence.}
L-PACT was designed as a gate-structured hierarchy rather than a single benchmark score (Fig.~\ref{fig:lpact_framework}A). Level 1 evaluates predictive adequacy by comparing real model features with nuisance baselines and the best severe control. Level 2 evaluates relational adequacy by comparing model-to-brain alignment profiles with brain-to-brain alignment patterns. Level 3 evaluates counterfactual mechanism-stripping adequacy by stripping candidate mechanisms and recomputing held-out neural scores within the model-based predictor. Level 4 evaluates reliability-bounded adequacy by normalizing model evidence against brain-brain ceilings. A positive integrated L-PACT candidate requires more than a predictive score: it must survive controls, reproduce relational structure, show mechanism-specific necessity, approach an available reliability bound, and replicate at the configured subject-run level.

This hierarchy prevents a local prediction score from carrying a stronger claim than it can support. Level 1 is predictive, Level 2 is relational, Level 3 is mechanism-stripping within the predictor, and Level 4 is reliability-bounded. Only a row that passes all required gates would justify the strongest candidate label. The current analysis does not produce such rows.

Formally, each candidate row \(r\) is represented by binary gate variables \(g_1(r),g_2(r),g_3(r),g_4(r)\) for predictive, relational, mechanism-stripping, and operational Turing-bounded reliability evidence, together with a severe-control indicator \(c(r)\) and a replication indicator \(\rho(r)\). The integrated L-PACT candidate indicator is the conjunctive decision rule
\begin{equation}
G_{\mathrm{LPACT}}(r)=g_1(r)g_2(r)g_3(r)g_4(r)c(r)\rho(r).
\label{eq:joint_gate}
\end{equation}
This product form is conjunctive: a row cannot compensate for an unmet mechanism-stripping or reliability gate by having a larger local predictive score. The final decision table is therefore a logical evidence matrix rather than a ranked model comparison, and control-explained rows remain informative outcomes.

This choice trades sensitivity for interpretability. L-PACT retains exploratory and diagnostic rows in the output tables, but the strongest labels require the conjunctive rule in Eq.~\ref{eq:joint_gate}. The framework is therefore designed to adjudicate strong claims rather than to increase the number of positive classifications.

\begin{figure}[htbp]
\centering
\includegraphics[width=\textwidth]{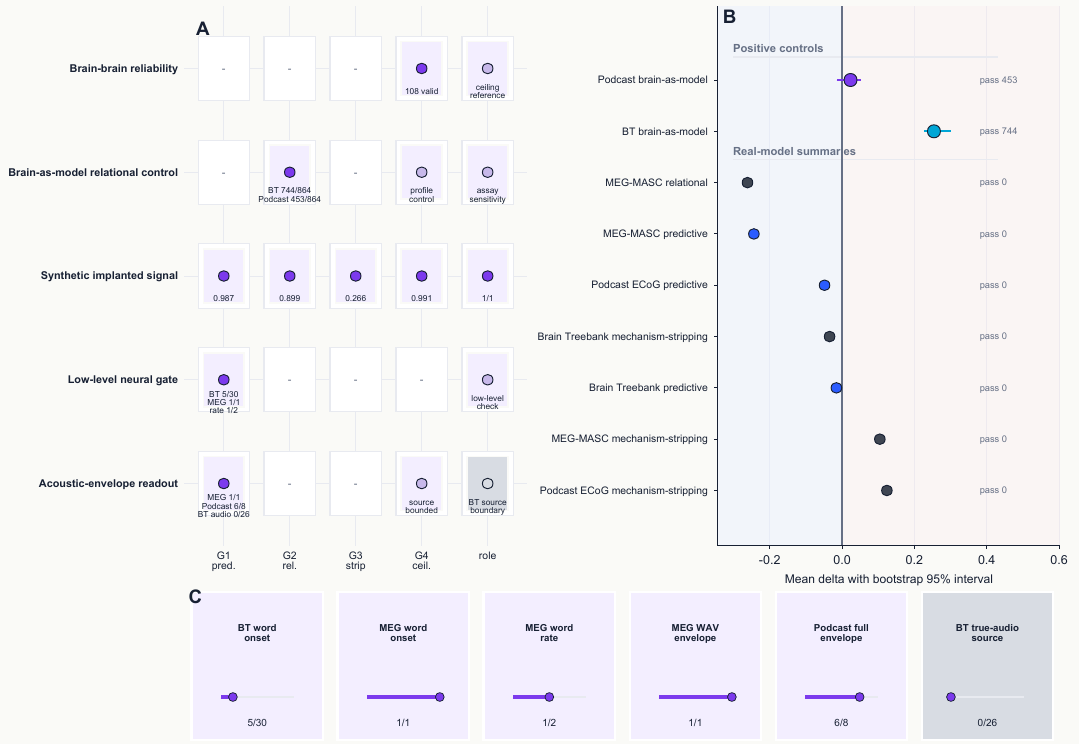}
\caption{Assay sensitivity and positive controls. (A) Positive-control gate matrix for brain-brain reliability, brain-as-model relational profiles, synthetic implanted signal, independent low-level neural gates, and standalone acoustic-envelope readouts. (B) Positive-control and real-model summary deltas with bootstrap 95\% intervals where available. (C) Low-level and acoustic assay-sensitivity checks, including the Brain Treebank true-audio source boundary. These controls show that L-PACT can detect stable brain-derived, low-level, acoustic, or known implanted structure without changing the real model decisions.}
\label{fig:assay_sensitivity}
\end{figure}

\textbf{Assay-sensitivity controls show that L-PACT can detect known structure.}
A fully negative model result would be difficult to interpret if the assay itself had no route to a positive outcome. We therefore evaluated positive-control checks without changing the real model decisions (Fig.~\ref{fig:assay_sensitivity}). These controls have different evidential roles. The synthetic implanted-signal control is an engineering check that the gate architecture can pass when known signal is present. Split-half brain reliability is a basic neural-artifact check: it can share low-frequency structure within a recording, so it is necessary but not by itself the strongest biological positive control. Brain-as-model run-to-run testing is stronger because one brain-derived profile is evaluated against another repeated brain profile rather than against a split of the same time series. Low-level word-onset, word-rate, and acoustic-envelope gates test whether the preprocessing can recover expected timing-linked neural signals.

The derived neural artifacts support valid brain-brain reliability estimates. All 108 valid brain-brain reliability rows exceed the configured minimum reliability of 0.1. Among primary datasets, split-half reliability is high for Brain Treebank, MEG-MASC, and Podcast ECoG, and Brain Treebank also has valid run-to-run reliability. These checks show that the brain side can produce stable positive structure even when model-to-brain rows do not pass.

Second, the brain-as-model relational control directly tests the profile machinery and is the strongest biological assay-sensitivity result in this study. A brain pattern from one run or subject is treated as the model-side profile and compared with the matching brain-unit profile from another run or subject, with shuffled brain-unit order as the control. Brain Treebank run-to-run profiles pass this L-PACT-like relational gate in all 9 alignment-by-profile metric summaries, covering 744 of 864 row-level comparisons. Podcast ECoG subject-to-subject profiles were available but did not pass the aggregate gate, providing a dataset-specific boundary case; by contrast, the stronger Brain Treebank repeated-run control passes, and the Podcast result records dataset-specific variability and coverage limits.

Third, independent low-level neural and acoustic-envelope checks show that the analysis chain can recover expected timing-linked signals where the required source media exist. Brain Treebank and MEG-MASC pass word-event sanity gates, and WAV-derived envelope controls pass for MEG-MASC and most readable full-length Podcast ECoG subject-runs. Brain Treebank is treated as a source-boundary case rather than relabeled as an acoustic-envelope positive control because the available public or local artifact chain lacks rights-cleared waveform or movie files. Finally, a deterministic implanted-signal simulation passes all four L-PACT gates. These controls do not provide empirical evidence that a language model aligns with brain data; they establish that L-PACT can detect brain-derived, low-level, acoustic, or implanted structure when the relevant signal is present.

\textbf{Source-audited dataset eligibility prevents inflated neural evidence.}
The source audit classifies datasets before evidence is interpreted (Fig.~\ref{fig:lpact_framework}B and C). Brain Treebank, Podcast ECoG, and MEG-MASC are the primary datasets because the analyzed artifact set contains derived brain time series, word events, model features, and enough matchable units for part of the L-PACT chain. Brain Treebank contributes 10 subjects, 26 subject-run units, 248 brain units, 11 matchable relational runs, and 11 matchable mechanism-stripping runs. Podcast ECoG contributes 9 subjects, 9 subject-run units, 235 brain units, 5 matchable relational runs, and 5 matchable mechanism-stripping runs. MEG-MASC contributes 11 subjects, 84 subject-run units, 257 brain units, 1 matchable relational run, and 18 matchable mechanism-stripping runs.

Narratives and LPP multilingual are retained as secondary or exploratory sources, and Learning Brain is validation-only. NaturalStories is treated as stimulus-LLM-only or diagnostic rather than primary neural evidence. This classification is inferentially consequential: including a dataset without a complete neural-side source chain would increase apparent evidence breadth without increasing evidential strength. The audit therefore functions as a scientific filter. It determines which datasets can support predictive analysis, relational profile analysis, mechanism stripping, and ceiling-bounded interpretation.

The audit is implemented as a dataset-level closure test. A confirmatory evidence chain requires brain time series, word/event grids, model features, severe controls, reliability estimates, and minimum coverage in subject-run and brain-unit counts. The displayed chain-complete flag, matchable relational-run count, and matchable mechanism-stripping-run count are direct operationalizations of this closure test in the feasibility table. Formal definitions and artifact field names are provided in the SI.

\begin{figure}[htbp]
\centering
\includegraphics[width=\textwidth]{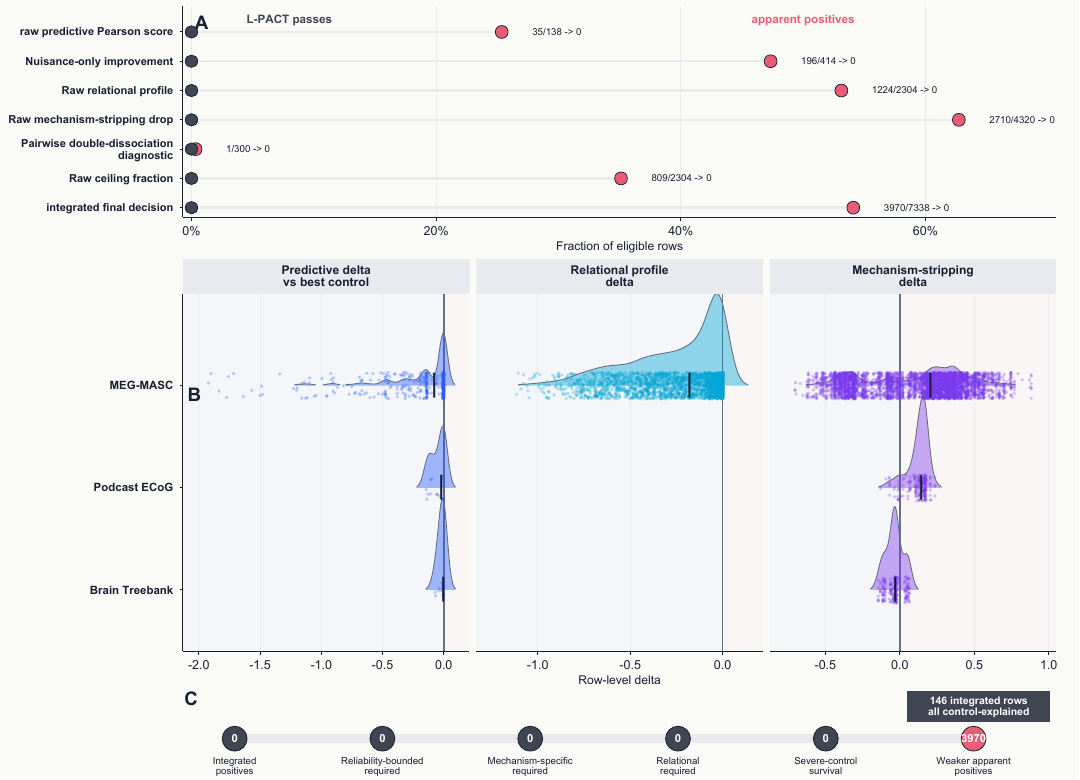}
\caption{Conventional-looking positives are downgraded by L-PACT. (A) Less stringent single-criterion rules count many apparent positives, whereas the corresponding L-PACT passing counts are zero. (B) Row-level predictive, relational, and mechanism-stripping delta distributions show that raw-looking signals do not become robust positive deltas over controls. (C) Progressive tightening of the evidential standard converts apparent positives into zero integrated positive rows, while all 146 integrated rows remain control-explained.}
\label{fig:conventional_downgrade}
\end{figure}

\textbf{Conventional-looking positives are downgraded by the L-PACT gates.}
The negative final decision is not due to an absence of all upstream signal. Less stringent single-criterion rules would have produced apparent positives (Fig.~\ref{fig:conventional_downgrade}). Thirty-five predictive-control rows have a positive raw Pearson model score, and 196 rows improve over the nuisance baseline if severe-control survival is ignored. In the relational stage, 1224 rows have positive raw model-to-brain profile similarity. In the mechanism-stripping stage, 2710 rows have positive raw stripping deltas, and 1 of 300 pairwise double-dissociation diagnostics is marked as passing. Finally, 809 ceiling-normalized rows reach a raw fraction of ceiling of at least 0.25 before the model-control delta is considered.

L-PACT rejects these conventional-looking positives because the stronger claim requires the appropriate contrast, not merely a favorable raw score. Predictive rows must beat nuisance and severe controls simultaneously. Relational rows must beat the best control profile. Mechanism-stripping rows must be mechanism-specific and supported by upstream gates. Ceiling-normalized rows must retain positive model-control deltas relative to a valid reliability estimate. Under these rules, the apparent positives are downgraded to control-explained evidence rather than promoted to alignment candidates.

\textbf{Uncertainty and coverage checks bound the negative result.}
The model-facing result is not driven by a single positive dataset being averaged away. Dataset-level bootstrap summaries over the implemented aggregation units show negative predictive deltas for Brain Treebank, Podcast ECoG, and MEG-MASC, and a negative relational profile delta for MEG-MASC (SI Tables 17--24). The available model-facing decision-ready deltas have limited independent subject-run support after full source, feature, control, and gate closure; one-unit intervals are therefore reported as coverage limitations rather than as strong population intervals.

The strongest biological positive control has a different uncertainty profile. Brain Treebank brain-as-model run-to-run profiles have a positive mean delta over shuffled unit order, \(0.254\) with bootstrap 95\% interval \(0.228\)--\(0.301\), over three source-target run pairs. Podcast ECoG subject-to-subject brain-as-model profiles are weaker, \(0.023\) with interval \(-0.015\)--\(0.053\), consistent with the manuscript's claim that this control is available but not aggregate-passing. A stricter brain-as-model oracle check shows the same boundary: Brain Treebank passes relational plus ceiling checks in 9 of 9 metric summaries, whereas Podcast ECoG passes 0 of 9; this control does not claim predictive or mechanism-stripping gates for brain-as-model rows.

Sensitivity analyses in the SI show that the conclusion is stable to reliability thresholds, fraction-of-ceiling thresholds, positive-direction thresholds, and best-control aggregation rules. The severe-control conclusion is not attributable to a single control family: removing any single severe-control family does not create an integrated L-PACT candidate, and weaker positive counts increase only under rules that ignore the corresponding contrast. Leave-one-dataset-out checks give the same final decision; omitting MEG-MASC leaves fewer integrated rows, but still yields no predictive, relational, mechanism-stripping, reliability-bounded, or integrated pass.

\textbf{Predictive evidence is control-explained under L-PACT gates.}
The Level 1 stage materializes 9036 held-out predictive-score rows and 414 decision-ready predictive-control delta rows: 18 deltas for Brain Treebank, 18 for Podcast ECoG, and 378 for MEG-MASC. Predictive adequacy requires that the real model score exceed both a nuisance baseline and the best severe control while using finite held-out scores, sufficient train/test samples, and subject-run aggregation. Severe controls include random matched-dimensionality features, autocorrelation-matched random features, circular shifts, sentence-reset features, reversed-context features, layer-label permutation, token-order shuffle, and within-story block shuffle. The most frequent best-control winners were circular-shifted model features, random matched-dimensionality features, autocorrelation-matched random features, and layer-label permutation.

For each predictive row, L-PACT compares the held-out real-model score against both the nuisance baseline and the strongest severe control. The predictive gate is positive only when these contrasts are positive under the configured finite-score, sample-size, diagnostic-status, and subject-run aggregation requirements. This formalization is applied directly to the locked predictive-control contrasts; exact contrast definitions are provided in the SI.

No integrated decision row passes the predictive gate. Brain Treebank and Podcast ECoG enter the final closure through matchable derived event, feature, and neural grids, but their mean deltas relative to the best severe controls are negative. MEG-MASC contributes the largest number of predictive rows, yet its apparent model effects are also control-explained under the gate. At the predictive-delta level, 218 rows are nuisance-explained and 196 rows are severe-control-explained; across the final decision table, the number of Level 1 passing rows is 0. This result supports a conservative conclusion: predictive evidence in the analyzed table set is insufficient for structural alignment.

\textbf{Relational alignment patterns do not support structural brain-language model alignment.}
Level 2 implements the alignment-pattern component of L-PACT. The stage materializes 768 brain-alignment-pattern rows and 216 model-alignment-pattern rows before profile comparison. Brain-to-brain profiles represent how each brain unit relates to other brain units, while model-to-brain profiles represent how a model layer or mechanism relates to the same ordered set of brain units. The relational test asks whether the model profile resembles the brain profile more than severe-control profiles do.

For each metric, the brain pattern is the vector of a brain unit's relations to other units, and the model pattern is the vector of a model layer or mechanism's relations to the same ordered brain-unit set. The decision-ready relational contrast compares the real profile similarity with the strongest comparable control profile using identical brain-unit ordering. The gate is therefore sensitive to the organization of the profile, not merely to the marginal magnitude of a model-to-brain score; formal vector definitions are provided in the SI.

The relational stage contains 20,736 raw profile-similarity score rows and 2304 delta-level rows. The raw table records real and control profile comparisons across metrics, whereas the delta table records the decision-ready contrast between each real profile and its best comparable control. These delta rows are concentrated in MEG-MASC, where the analyzed derived artifacts provide a matchable model-feature and brain-unit grid. No row passes the relational gate. Mean real profile similarity is positive but small, the mean delta relative to the best control is negative, and ceiling-normalized evidence remains far below the required gate.

The relational result is informative precisely because it tests a limitation that ordinary encoding scores cannot isolate. A model can predict many units at a weak but nonzero level while still imposing the wrong ordering over units, regions, or mechanisms. Conversely, a relational profile could in principle be meaningful even when individual unit scores are modest, provided that the cross-unit pattern exceeds controls and approaches a valid brain-brain ceiling. In the locked outputs, the first condition is not met: every relational delta row is control-explained, with token-order shuffle, circular shift, autocorrelation-matched random features, and layer-label permutation frequently serving as the strongest controls.

\textbf{Mechanism stripping does not yield mechanism-specific candidates.}
Level 3 evaluates whether candidate model mechanisms are necessary in a mechanism-specific way within the implemented feature-readout analysis. The mechanism-stripping stage materializes 720 stripping-feature metadata rows, 8640 real-or-stripped score rows, 4320 stripping-delta rows, and 300 double-dissociation tests. The stripping-delta rows include 216 for Brain Treebank, 216 for Podcast ECoG, and 3,888 for MEG-MASC. Mechanisms include surprisal, semantic transition, dependency integration, syntactic boundary, discourse boundary, and context update. Stripping methods include feature zeroing, feature residualization, mechanism-specific projection removal, layer ablation, context reset, and reversed context. Where derived grids are matchable, held-out stripped scores are recomputed with blocked train/test splits and train-only inner selection over lag, ridge alpha, and PCA dimensionality.

For each stripped mechanism and neural target mechanism, the implemented contrast is the held-out score drop after stripping. Specificity is measured by comparing the matching drop with nonmatching target drops. Thus, a nonspecific degradation of feature quality is not sufficient: the matching mechanism must show a larger drop than nonmatching mechanisms under the same cross-validation and control logic. Exact formulas are reported in the SI.

No row passes the Level 3 gate. Some stripping deltas are positive, and one of 300 double-dissociation tests is marked as a pairwise diagnostic pass, but this diagnostic does not satisfy the integrated L-PACT gate. A mechanism-specific candidate must also be supported by predictive and relational gates, survive severe controls, respect subject-run aggregation, and show selectivity. Because those conditions are not met, all 4320 stripping-delta rows are recorded with the upstream reason \texttt{no\_predictive\_candidate}, and the integrated candidate table contains 0 rows. This rules out mechanism-equivalence claims under the current data and gates without implying that future models or richer neural datasets could never satisfy the framework.

The stripping analyses test necessity of representational components for held-out prediction within the implemented feature and readout analysis; they do not intervene on a biological system. Under the current gates, observed drops are nonspecific, unsupported by upstream evidence levels, or insufficiently robust to controls for an integrated L-PACT decision.

\textbf{Reliability-bounded evidence does not pass the operational Turing-bounded gate.}
Level 4 estimates brain-brain ceilings with split-half, run-to-run, subject-to-subject, and session-to-session logic where the derived artifacts contain the required repeated-measure structure. The ceiling table contains 420 rows, of which 108 are valid under the configured reliability checks. Split-half ceilings are available for Brain Treebank, Podcast ECoG, MEG-MASC, and LPP multilingual; run-to-run estimates are valid for Brain Treebank; other modes are explicitly recorded as invalid when the required subject, run, or session pairing is absent. No final decision row receives the operational Turing-bounded reliability pass.

The Level 4 score is explicitly reliability-normalized: model scores and model-control deltas are divided by the valid brain-brain reliability estimate for the corresponding dataset and metric. Rows without a valid reliability estimate, or with reliability below the configured threshold, cannot receive the operational Turing-bounded reliability pass. In the decision table, the number of Level 4 passing rows is 0, reinforcing that the integrated outcome is not driven by an unnormalized local score.

\begin{figure}[htbp]
\centering
\includegraphics[width=\textwidth]{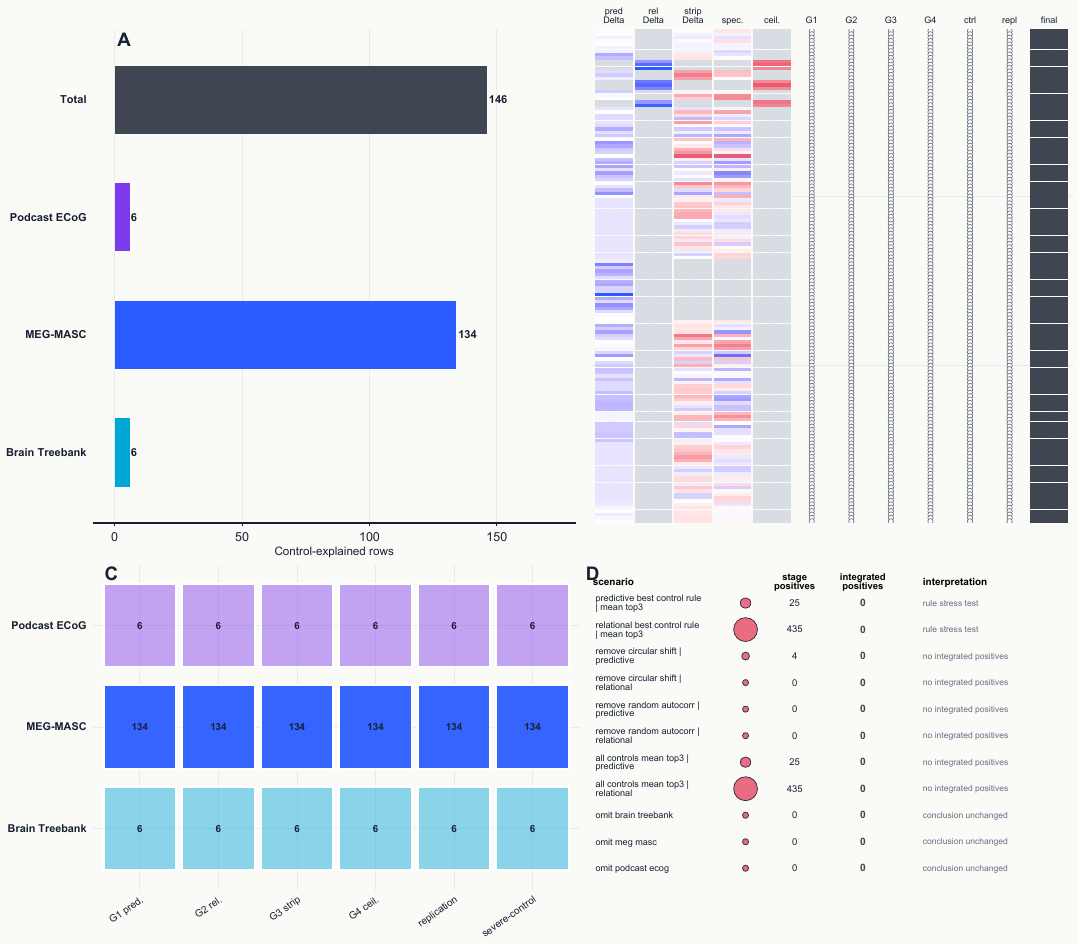}
\caption{Final integrated decision, nonpassing taxonomy, and robustness. (A) All 146 integrated rows are control-explained, with 6 Brain Treebank rows, 134 MEG-MASC rows, and 6 Podcast ECoG rows. (B) The recorded row-level nonpassing label is severe-control explained for every integrated row. (C) Key threshold-sensitivity, control-family ablation, and leave-one-dataset-out scenarios show stage-level apparent positives where present, but no integrated positive rows.}
\label{fig:final_decision}
\end{figure}

\textbf{Final L-PACT decision: apparent correspondence is control-explained.}
The final decision table integrates predictive, relational, mechanism-stripping, ceiling, severe-control, and replication gates (Fig.~\ref{fig:final_decision}). All 146 rows are classified as control-explained. This total includes 134 MEG-MASC rows and 12 additional closure rows from Brain Treebank and Podcast ECoG. There are no predictive-only rows, no relational-candidate rows, no mechanism-specific candidate rows, and no integrated positive candidates. The final decision is Outcome D: no robust predictive or structural alignment under current controls.

The supplementary material makes this negative-control conclusion auditable. SI Tables 1--26 record dataset eligibility, gate definitions, controls, assay-sensitivity checks, Brain Treebank acoustic-proxy boundaries, conventional-vs-LPACT contrasts, uncertainty summaries, threshold sensitivity, control-family ablation, brain-as-model oracle checks, leave-one-out stability, model inventory, decision summaries, scope and limitation summaries, mechanism-stripping diagnostics, and the index of row-level CSV files. Reproducibility artifacts are documented in the SI.

\section*{Discussion}

The central result is not that language models and brains are unrelated. It is that prediction-like scores, even when numerous and apparently favorable, are not sufficient evidence of structural or mechanism-specific alignment once source audit, severe controls, relational profiles, mechanism stripping, and brain-brain reliability are imposed. In the analyzed derived artifact set, no integrated row passes the gates required for positive structural or mechanism-specific evidence. The result should therefore be read as a constraint on evidential strength, not as a universal claim about every possible model-brain relationship.

This distinction matters for NeuroAI because high predictive scores can arise from nonspecific structure. Naturalistic language stimuli contain temporal autocorrelation, lexical regularities, contextual dependencies, and dimensional properties that can help flexible readouts predict neural measurements. Severe controls are therefore not ancillary diagnostics; they are necessary for deciding whether an apparent model advantage is specific to the real representation. In this study, the best severe controls explain the integrated evidence.

The findings also connect language-model evaluation to critiques from visual NeuroAI and brain-model benchmarking. Single-score comparisons may lack discriminative power when different models or controls produce similar local scores \cite{cadieu2014,khaligh2014,yamins2016,kell2018,richards2019,apa,algonauts}. Relational tests ask whether a model reproduces the organization among neural units rather than merely matching a single target. In the analyzed data, the relational alignment-pattern test does not support a positive structural claim, and ceiling-normalized evidence remains below the required standard.

The conclusion can coexist with prior reports of language-model and neural correspondence. Many previous studies ask whether model features predict neural responses or whether model layers track aspects of language processing. L-PACT asks a stricter question: whether those correspondences survive controls, reproduce relational structure, show mechanism-specific necessity, and approach brain-brain reliability. A weaker positive result in a predictive analysis is therefore not contradicted by a negative L-PACT decision; it is assigned to a different evidence level.

Several limitations bound the conclusion. The analysis depends on available derived artifacts and their preprocessing. Dataset coverage is uneven across evidence levels: relational profile deltas are concentrated in MEG-MASC, while mechanism-stripping closure is available for Brain Treebank and Podcast ECoG through matchable derived grids. Model coverage is a substantive boundary: the negative result applies to the locked model-feature stores and controls tested here, not to the full space of contemporary or future language models. The conclusion should not be read as a model comparison benchmark for the full space of contemporary LLMs. The decision table includes DistilGPT-2, GPT-2, GPT-2 Medium, Pythia-160M, Qwen2.5-1.5B-Instruct, and Qwen3-1.7B, while larger or missing registry models either exceed the study inclusion criteria or lack validated feature rows. This boundary does not exclude the possibility that larger, instruction-tuned, or otherwise different models could satisfy L-PACT in future data. Controls are severe but not exhaustive. Brain-brain ceiling estimates depend on the reliability structure available in the derived data, and some reliability modes are unavailable for some datasets.

The framework is also conservative by design, and that conservatism has a cost. Severe controls can absorb variance that partly overlaps with genuine language-related structure, and a conjunctive gate can convert partial evidence into a negative final label. This conservatism reflects the intended use of the framework: adjudicating strong structural and mechanism-specific claims. It also means that L-PACT should be read alongside exploratory tables rather than as a device for discarding all subthreshold observations. The strongest use of the framework is to distinguish claim strength, not to deny that weaker correspondences can motivate follow-up experiments.

Future work should test larger, instruction-tuned, and more diverse language models, prospective neural datasets, richer language-event annotations, and repeated subject-run designs that support stronger ceiling estimates. Source-localized MEG, broader ECoG coverage, and preregistered mechanism-specific manipulations would make the mechanism-stripping and relational tests more decisive. The broader contribution of L-PACT is methodological: it turns negative-control results into an explicit standard for future brain-language model comparison.

\section*{Materials and Methods}
\textit{L-PACT framework and source audit.} L-PACT is a four-level framework for evaluating brain-language model correspondence. Level 1 tests predictive adequacy; Level 2 tests relational adequacy; Level 3 tests counterfactual mechanism-stripping adequacy within the model-based predictor; and Level 4 tests reliability-bounded adequacy relative to brain-brain estimates. Positive structural or mechanism-specific evidence requires the required gates to pass jointly. Rows in which severe controls explain the effect are retained as scientific results rather than discarded.

The analysis was performed on locked derived artifacts rather than on redistributed raw neural files. The source audit classified Brain Treebank, Podcast ECoG, and MEG-MASC as primary datasets, retained Narratives and LPP multilingual outside the primary confirmatory chain, treated Learning Brain as validation-only, and excluded NaturalStories from primary neural evidence because the available artifact chain did not satisfy the configured neural-side requirements. Confirmatory interpretation required subject-run aggregation, severe controls, and an available brain-brain ceiling. The audit follows reproducibility principles that separate provenance and preprocessing status from downstream inferential claims \cite{nastase2021,gorgolewski2016,markiewicz2021,gramfort2013,poldrack2017}.

\textit{Model representations and predictive readout.} The analysis uses frozen language-model feature stores already present in the locked analysis set; no new model extraction was performed for this manuscript. Predictive adequacy compares real model scores with nuisance baselines and severe controls. Where held-out scores are recomputed from matchable derived grids, neural prediction uses ridge-regularized linear encoding within blocked cross-validation \cite{hoerl1970,hastie2009,pedregosa2011}. Lag, ridge penalty, PCA dimensionality, residualization, and projection-removal steps are selected or fit only inside training data, following standard leakage-prevention cautions \cite{stone1974,varma2006,varoquaux2017,yarkoni2017}.

\textit{Relational, mechanism-stripping, and reliability-bounded stages.} Relational analysis builds brain-to-brain and model-to-brain profiles over a shared brain-unit order and compares profile similarity against severe controls using correlation, cosine, RSA-style, event-response, and centered-kernel alignment metrics \cite{mantel1967,haxby2001,kriegeskorte2008,cortes2012,gretton2005,kornblith2019}. Mechanism stripping tests surprisal, semantic transition, dependency integration, syntactic boundary, discourse boundary, and context update by recomputing held-out scores after feature removal, residualization, projection removal, layer ablation, context reset, or reversed context. Double-dissociation tests are diagnostic and are not sufficient by themselves for an integrated L-PACT decision. Brain-brain ceilings are estimated from split-half, run-to-run, subject-to-subject, and session-to-session reliability where the derived artifacts support those comparisons; invalid or subthreshold ceilings cannot be promoted to reliability-bounded evidence.

\textit{Assay-sensitivity controls.} Positive-control checks do not alter real model decisions. They include brain-brain reliability, brain-as-model relational profiles, profile-order implementation controls, independent low-level neural gates, standalone acoustic-envelope readouts where source media are available, Brain Treebank source-boundary audits for true audio, and a deterministic implanted-signal simulation. These controls show that the analysis can detect stable brain-derived, low-level, acoustic, or known implanted structure while preserving nonpassing real-model outcomes.

\textit{Statistical inference and claim audit.} The final decision table integrates predictive, relational, mechanism-stripping, reliability-bounded, severe-control, and replication gates. Dataset-level uncertainty summaries aggregate over the implemented subject-run or source-target units before bootstrapping; one-unit intervals after full gate closure are reported as coverage limitations. Where inferential summaries are reported, resampling, permutation-style diagnostics, and false-discovery-rate correction follow standard bootstrap, nonparametric neuroimaging, and Benjamini-Hochberg logic \cite{efron1994,nichols2002,benjamini1995}. The SI gives implementation equations, row-level table maps, sensitivity analyses, and scope and limitation summaries.

\section*{Data, Materials, and Software Availability}
Derived result tables, analysis code, manuscript tables, and reproducibility scripts are available through an OSF view-only review link: \url{https://osf.io/ab27g/overview?view_only=2ab1722284b243bca177ebdde4b655f3}. Raw neural datasets and stimulus media are not redistributed and must be obtained from their original public repositories or data owners under the terms set by those providers. Private API/model scratchpads and credentials are excluded from all public artifacts. Upon acceptance, the repository record will be made public and assigned or updated with a persistent accession or DOI before publication.

\section*{Acknowledgments}
The author received no specific funding for this work. The author thanks the investigators, participants, and data stewards who made the public neural and stimulus-derived datasets available, and the maintainers of the open-source scientific Python, neuroimaging, data-processing, and LaTeX tools used in this work.

\clearpage
\appendix
\setcounter{figure}{0}
\setcounter{table}{0}
\renewcommand{\thefigure}{S\arabic{figure}}
\renewcommand{\thetable}{S\arabic{table}}
\section*{Supplementary Information}
\subsection*{Supplementary Overview}

This Supplementary Information (SI) documents the implementation-faithful evidence map for the main \lpact{} manuscript. It expands the main Materials and Methods without changing any scientific result, gate decision, or row count. All values reported here are taken from the locked derived \lpact{} artifacts and exported manuscript tables. Raw neural data and stimulus media are not redistributed, moved, or modified by this SI.

\lpact{} stands for Language Predictive, Alignment-pattern, Causal, and Turing-bounded Test. The acronym retains ``Causal'' and ``Turing-bounded'' as part of the established L-PACT name, but the implemented gates are described here as counterfactual mechanism-stripping and reliability-bounded interpretation. Mechanism stripping is a model-feature intervention within the implemented predictor, not a biological intervention; reliability-bounded interpretation means comparison with available brain-brain estimates, not a claim that the NeuroAI Turing Test is a settled field standard. The framework separates four evidence levels that are often conflated in brain-language model comparisons:

\begin{enumerate}
  \item Level 1, predictive adequacy: model-derived features must improve held-out neural prediction relative to nuisance baselines and the strongest available severe control.
  \item Level 2, relational adequacy: model-to-brain alignment profiles must reproduce brain-to-brain alignment profiles over a shared brain-unit order.
  \item Level 3, counterfactual mechanism-stripping adequacy: removing a candidate mechanism from model features must selectively damage prediction for matching neural targets more than for nonmatching targets within the implemented predictor.
  \item Level 4, reliability-bounded adequacy: the surviving model evidence must be interpreted relative to brain-brain reliability or ceiling estimates.
\end{enumerate}

The locked analysis package reports a control-explained model outcome. The final decision table contains 146 integrated rows. All 146 rows are labelled \code{control_explained}. No row passes the predictive, relational, mechanism-stripping, or reliability-bounded gate, and no row qualified as an integrated L-PACT candidate. This SI records the table-level basis for these bounded interpretations.

The SI includes assay-sensitivity, conventional-contrast, sensitivity, ablation, coverage, and scope-boundary tables. Brain-brain reliability controls and a deterministic synthetic implanted-signal control show that the L-PACT decision architecture can produce positive outcomes when stable brain-derived or known implanted structure is present. The conventional-contrast table shows that less stringent raw-score rules would have counted apparent positives that the L-PACT gates downgrade to control-explained evidence.

\section*{Relationship to the Main Manuscript}

The main manuscript reports the conceptual claim and core evidence in a compressed form. This SI provides the detailed artifact accounting needed to audit that claim. In particular, it records:

\begin{itemize}
  \item dataset roles and eligibility decisions;
  \item evidence-chain closure across predictive, relational, mechanism-stripping, and ceiling stages;
  \item gate definitions and the decision rule used for final labels;
  \item severe controls, nonpassing labels, and claim boundaries;
  \item assay-sensitivity positive controls and conventional-vs-LPACT contrasts;
  \item bootstrap uncertainty, threshold sensitivity, control-family ablation, brain-as-model oracle checks, and leave-one-out summaries;
  \item model inventory and model-coverage boundaries;
  \item locked artifact row counts;
  \item mechanism-stripping diagnostic tables;
  \item scope and limitation summaries for common boundary and conservatism questions.
\end{itemize}

The SI does not introduce new model extraction, new neural preprocessing, new gate thresholds, or new exploratory positives. The manuscript and SI should be interpreted against the same locked analysis package.

\section{Supplementary Figures}

\begin{figure}[p]
\centering
\includegraphics[width=\textwidth]{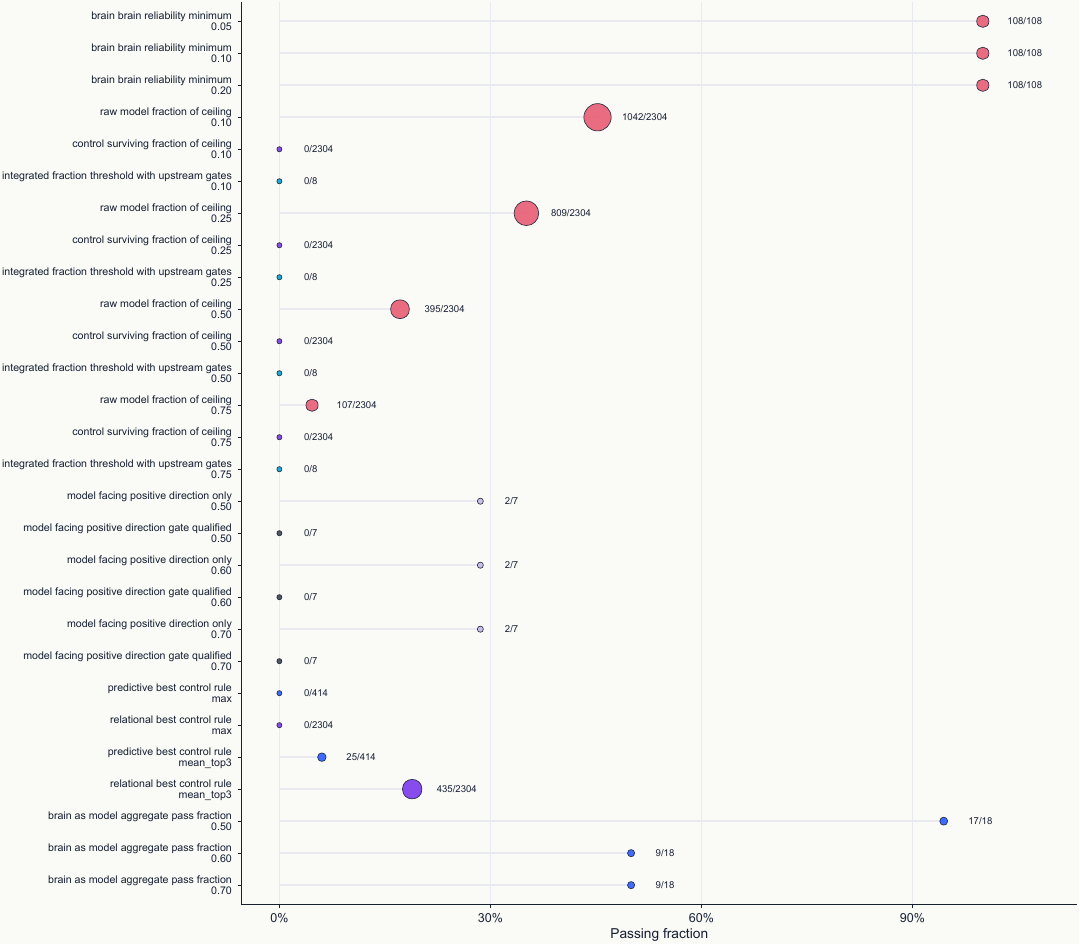}
\caption{Full gate-threshold sensitivity matrix. The matrix reports passing rows, eligible rows, and passing fractions for reliability thresholds, fraction-of-ceiling thresholds, positive-direction thresholds, and best-control aggregation rules. No threshold or rule family creates integrated positive rows.}
\label{fig:s1_threshold_sensitivity}
\end{figure}

\begin{figure}[p]
\centering
\includegraphics[width=\textwidth]{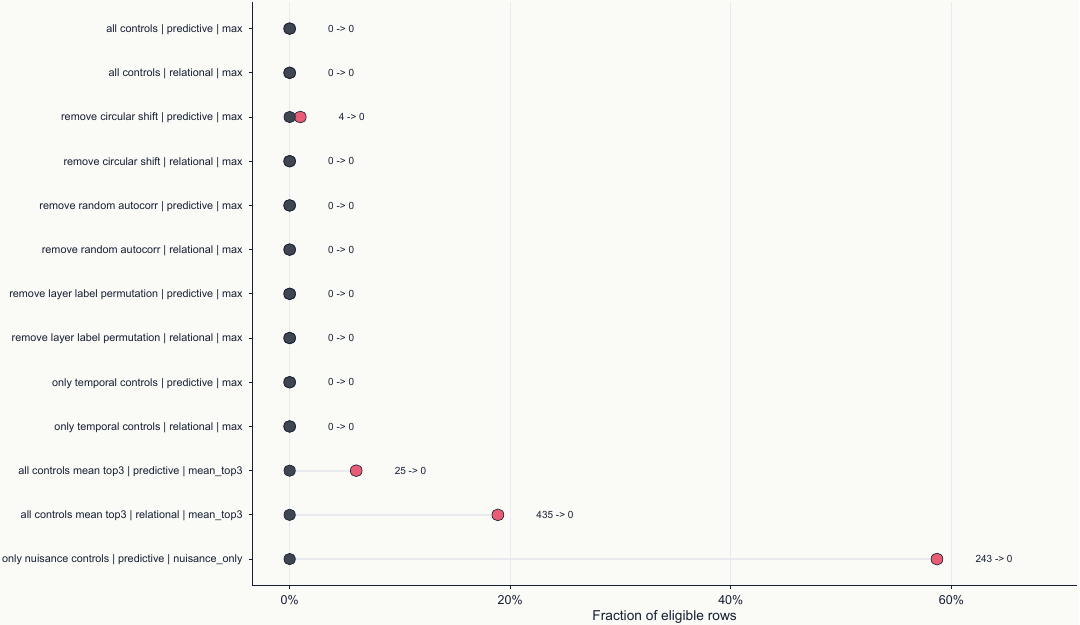}
\caption{Control-family ablation summary. Removing or weakening severe-control families can increase stage-level apparent positive rows under selected rules, but integrated positive rows remain zero.}
\label{fig:s2_control_family_ablation}
\end{figure}

\begin{figure}[p]
\centering
\includegraphics[width=\textwidth]{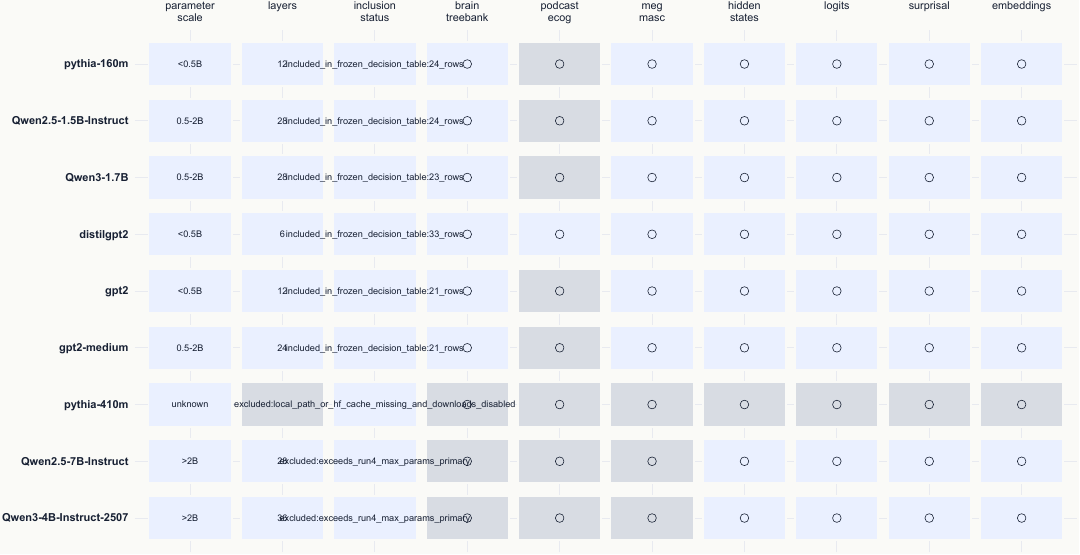}
\caption{Model inventory and coverage heatmap. Rows are tested or registry-listed model entries; columns summarize family, parameter scale, layer count, dataset coverage, feature families, available controls, row availability, and inclusion status.}
\label{fig:s3_model_inventory}
\end{figure}

\begin{figure}[p]
\centering
\includegraphics[height=0.88\textheight,width=\textwidth,keepaspectratio]{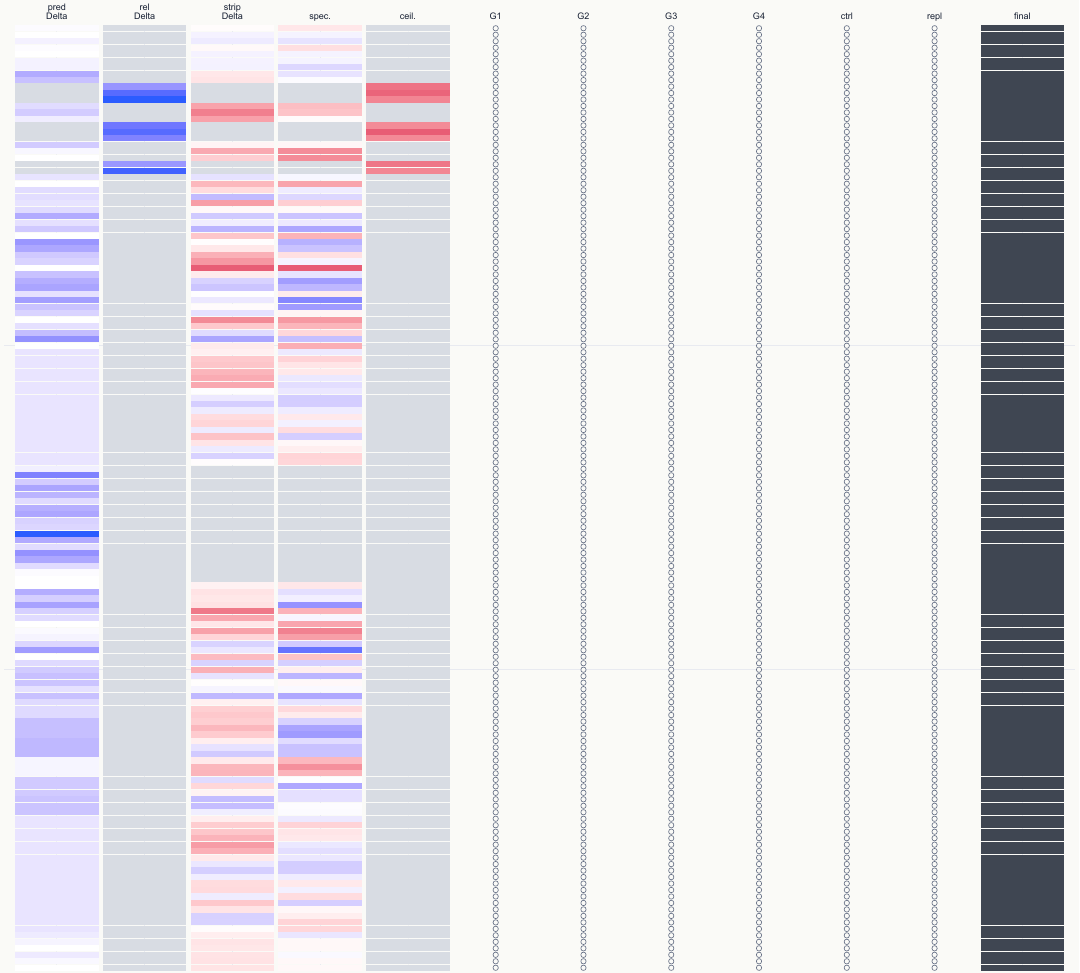}
\caption{Row-level integrated decision heatmap. Rows are the 146 integrated decision rows, sorted by dataset, mechanism, model, region group, and time window. Numeric columns are scaled within column for visualization; outcome columns show the final decision and recorded failure reason.}
\label{fig:s4_row_level_decision}
\end{figure}

\begin{figure}[p]
\centering
\includegraphics[width=\textwidth]{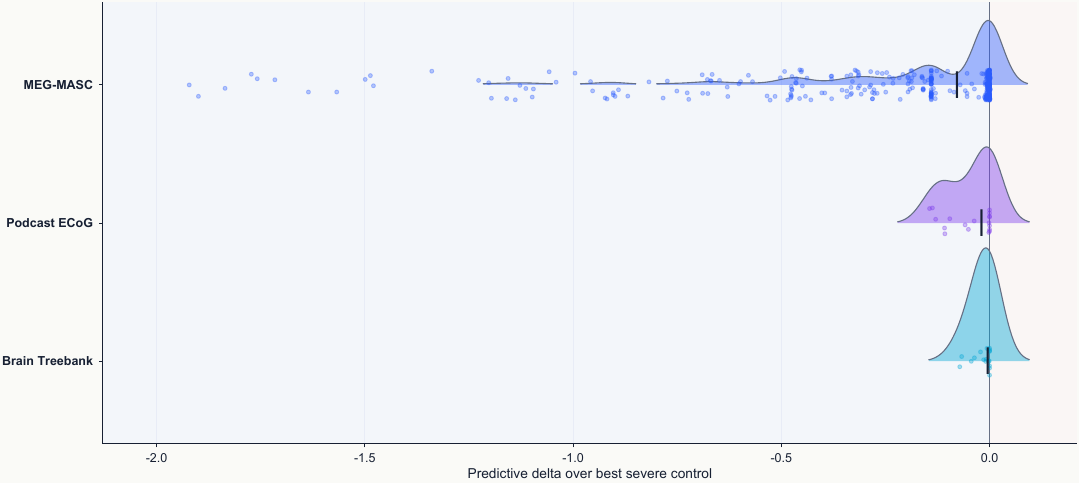}
\caption{Predictive row-level delta distributions. The figure summarizes predictive deltas over the best severe control and nuisance baseline by dataset using the locked row-level predictive-control table.}
\label{fig:s5_predictive_deltas}
\end{figure}

\begin{figure}[p]
\centering
\includegraphics[width=\textwidth]{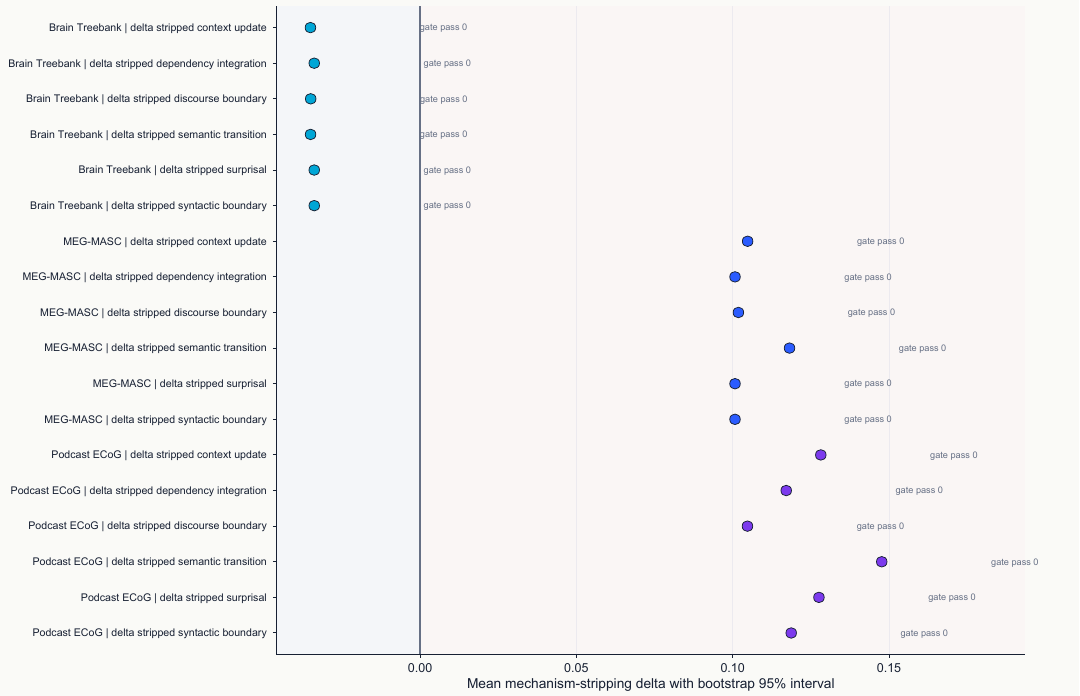}
\caption{Mechanism-stripping uncertainty by stripped mechanism. Points show mean mechanism-stripping deltas with bootstrap 95\% intervals where available. Positive stripping deltas are not interpreted as mechanism-specific candidates unless upstream gates also pass.}
\label{fig:s6_mechanism_stripping_uncertainty}
\end{figure}

\clearpage

\section{Source Audit and Dataset Eligibility}

Datasets are classified before evidence is interpreted. This is a central design choice in \lpact{} because evidence strength depends on the available source chain, not only on the number of stimuli or reported subjects. A dataset can have useful stimulus or model features while still being unsuitable for primary neural evidence if the artifact chain lacks matchable neural time series, word events, controls, or brain-brain reliability structure.

The primary datasets in this manuscript are Brain Treebank, Podcast ECoG, and MEG-MASC. Narratives and LPP multilingual are retained outside the primary confirmatory neural chain. Learning Brain is validation-only. NaturalStories is treated as stimulus-LLM-only or diagnostic because the available artifact chain does not satisfy the configured neural-side evidence requirements for primary inference.

\begin{landscape}
\begingroup
\small
\setlength{\tabcolsep}{3pt}
\begin{longtable}{L{2.6cm} L{3.2cm} C{1.3cm} C{1.1cm} C{1.3cm} C{1.8cm} C{1.5cm} C{1.2cm} C{1.5cm} L{2.4cm}}
\caption{Dataset inventory and eligibility summary. The table reports the source-audit fields used by the PNAS manuscript. Subject-run units, brain units, model availability, and mechanism coverage are derived from the locked feasibility table. NaturalStories counts refer to derived diagnostic artifacts in the locked analysis package, not to a source-complete primary neural evidence chain; these rows are excluded from primary L-PACT inference. Brain units denote the units available in the derived artifact chain for L-PACT construction; they are not treated as independent subjects for inference.}\\
\toprule
Dataset & Role & Subjects & Runs & Stimuli & Subject-run units & Brain units & Models & Mech. & Feasibility status \\
\midrule
\endfirsthead
\toprule
Dataset & Role & Subjects & Runs & Stimuli & Subject-run units & Brain units & Models & Mech. & Feasibility status \\
\midrule
\endhead
\code{brain_treebank} & primary & 10 & 7 & 33 & 26 & 248 & 6 & 8 & confirmatory \\
\code{podcast_ecog} & primary & 9 & 9 & 10 & 9 & 235 & 1 & 8 & confirmatory \\
\code{meg_masc} & primary & 11 & 84 & 1344 & 84 & 257 & 6 & 9 & confirmatory \\
\code{narratives} & secondary & 2 & 3 & 3 & 6 & 6 & 1 & 0 & diagnostic \\
\code{lpp_multilingual} & secondary & 1 & 1 & 28 & 1 & 101 & 1 & 0 & exploratory \\
\code{naturalstories} & stimulus-LLM-only or diagnostic & 78 & 8 & 18 & 205 & 26 & 5 & 0 & diagnostic \\
\code{learning_brain} & validation-only & 12 & 0 & 0 & 92 & 101 & 0 & 0 & diagnostic \\
\bottomrule
\end{longtable}
\endgroup
\end{landscape}

\begin{landscape}
\begingroup
\scriptsize
\setlength{\tabcolsep}{2pt}
\begin{longtable}{L{2.6cm} C{0.85cm} C{0.85cm} C{0.85cm} C{0.85cm} C{1.15cm} C{1.05cm} C{0.9cm} C{1.15cm} C{0.85cm} C{1.05cm} C{1.45cm} L{1.7cm}}
\caption{Evidence-chain closure and feasibility flags. A confirmatory interpretation requires the configured source chain, subject-run aggregation, severe controls, and an available brain-brain ceiling. Rows that cannot satisfy the chain remain exploratory or diagnostic.}\\
\toprule
Dataset & Brain & Words & Model & Ctl. & Ceiling possible & Pred. & Rel. & Mech.-strip & Reliab. & Rel. runs & Mech.-strip runs & Evidence complete \\
\midrule
\endfirsthead
\toprule
Dataset & Brain & Words & Model & Ctl. & Ceiling possible & Pred. & Rel. & Mech.-strip & Reliab. & Rel. runs & Mech.-strip runs & Evidence complete \\
\midrule
\endhead
\code{brain_treebank} & \yes & \yes & \yes & \yes & \yes & \yes & \yes & \yes & \yes & 11 & 11 & \yes \\
\code{podcast_ecog} & \yes & \yes & \yes & \yes & \yes & \yes & \yes & \yes & \yes & 5 & 5 & \yes \\
\code{meg_masc} & \yes & \yes & \yes & \yes & \yes & \yes & \yes & \yes & \yes & 1 & 18 & \yes \\
\code{narratives} & \yes & \yes & \yes & \no & \yes & \no & \no & \no & \yes & 0 & 0 & \no \\
\code{lpp_multilingual} & \yes & \yes & \yes & \no & \yes & \no & \yes & \no & \yes & 0 & 0 & \no \\
\code{naturalstories} & \yes & \yes & \yes & \no & \no & \no & \no & \no & \no & 0 & 0 & \no \\
\code{learning_brain} & \yes & \no & \no & \no & \yes & \no & \no & \no & \yes & 0 & 0 & \no \\
\bottomrule
\end{longtable}
\endgroup
\end{landscape}

\subsection*{NaturalStories Boundary}

NaturalStories is not used as primary neural evidence in this manuscript. This exclusion is not a negative judgment about the dataset in general. It is a study-specific boundary: the available derived artifact chain does not satisfy the configured neural-side evidence requirements for primary \lpact{} inference. Retaining NaturalStories as diagnostic preserves transparency while preventing source breadth from being mistaken for evidential strength.

\section{Evidence Levels and Decision Gates}

\lpact{} uses conjunctive gates rather than a compensatory score. A row with a large isolated score cannot compensate for an unmet relational, mechanism-stripping, or reliability-bounded gate. This is the main reason the analysis can contain many derived rows while still producing no positive integrated candidates.

For each candidate row \(r\), the final integrated decision depends on predictive, relational, mechanism-stripping, and reliability-bounded gate variables \(g_1(r)\), \(g_2(r)\), \(g_3(r)\), and \(g_4(r)\), plus a severe-control indicator \(c(r)\) and a replication indicator \(\rho(r)\):

\[
I_{\mathrm{LPACT}}(r) =
g_1(r)g_2(r)g_3(r)g_4(r)c(r)\rho(r).
\]

The product form is deliberately conservative. It makes a strong interpretation possible only when every required evidence level is satisfied.

\begin{longtable}{L{3.2cm} L{11.0cm}}
\caption{L-PACT gate definitions and interpretation.}\\
\toprule
Gate & Operational definition \\
\midrule
\endfirsthead
\toprule
Gate & Operational definition \\
\midrule
\endhead
Predictive adequacy & The real model score must exceed nuisance baselines and the best severe control, with finite held-out scores, sufficient train/test samples, and subject-run aggregation before inference. \\
Relational adequacy & The model-to-brain profile must exceed the best control profile, use the same brain-unit ordering as the brain profile, and satisfy brain-pattern reliability constraints. \\
Counterfactual mechanism-stripping adequacy & Mechanism stripping must produce a positive and selective matching deficit within the implemented predictor. The matching drop must exceed nonmatching drops, survive controls, and not be reducible to generic feature degradation. \\
Reliability-bounded adequacy & Surviving model evidence must reach the configured fraction of a valid brain-brain reliability estimate. Invalid or missing ceiling estimates cannot be silently ignored. \\
Severe-control survival & The real model or real profile must beat random matched-dimensionality, autocorrelation-matched, temporal-shift, context-reset, token-order, block-shuffle, layer-permutation, or mechanism-mismatch controls where applicable. \\
Replication & The positive direction must hold across the configured fraction of subject-run units and satisfy the minimum unit requirements for a confirmatory claim. \\
\bottomrule
\end{longtable}

\section{Formal Contrasts and Implementation-Faithful Evidence Mapping}

This section contains the formal details that are intentionally compressed in the main Results. The main text retains the top-level conjunctive gate, while the SI records the operational contrasts used by the locked tables.

For dataset \(d\), let \(B_d,W_d,F_d,C_d\), and \(R_d\) denote availability of brain time series, word/event grids, model features, severe controls, and reliability estimates. Let \(U_d\) denote the configured minimum coverage in subject-run units and brain units. L-PACT treats the confirmatory evidence chain as complete only when
\[
E_d = B_d \land W_d \land F_d \land C_d \land R_d \land U_d .
\]

For each predictive row \(r\), the implemented predictive contrasts are
\[
\Delta_{\mathrm{nuis}}(r)=s_{\mathrm{real}}(r)-s_{\mathrm{nuis}}(r),\quad
\Delta_{\mathrm{ctrl}}(r)=s_{\mathrm{real}}(r)-\max_{c\in\mathcal{C}}s_c(r),
\]
\[
\Delta_{\mathrm{all}}(r)=s_{\mathrm{real}}(r)-\max\left(s_{\mathrm{nuis}}(r),\max_{c\in\mathcal{C}}s_c(r)\right),
\]
where \(\mathcal{C}\) is the severe-control set. A predictive row cannot pass when the best severe control or nuisance baseline matches or exceeds the real-model score.

For relational profiles, let \(u_1,\ldots,u_N\) denote ordered brain units on a canonical event grid and \(a_m(\cdot,\cdot)\) the implemented alignment score for metric \(m\). The brain pattern for unit \(u_i\) is
\[
\mathbf{b}_i^m=\left[a_m(u_i,u_j)\right]_{j\ne i},
\]
and the model pattern for model \(q\), layer \(\ell\), and mechanism \(h\) is
\[
\mathbf{p}_{q,\ell,h}^m=\left[a_m(z_{q,\ell,h},u_j)\right]_{j=1}^N.
\]
For profile-similarity metric \(S\), the decision-ready relational contrast is
\[
\Delta_{\mathrm{rel}}=S(\mathbf{p}_{\mathrm{real}},\mathbf{b})-\max_{c\in\mathcal{C}}S(\mathbf{p}_c,\mathbf{b}),
\]
with identical brain-unit order enforced for real and control profiles.

For stripped mechanism \(M\) and neural target mechanism \(T\), the stripping contrast is
\[
\Delta_{\mathrm{strip}}(M,T)=s_{\mathrm{real}}(T)-s_{\mathrm{stripped}}(M,T).
\]
Mechanism specificity requires this matching deficit to exceed nonmatching deficits:
\[
\eta_{M,T}=\Delta_{\mathrm{strip}}(M,T)-\frac{1}{|\mathcal{T}\setminus\{T\}|}\sum_{T'\in\mathcal{T}\setminus\{T\}}\Delta_{\mathrm{strip}}(M,T').
\]

For Level 4, if \(R_{BB}(d,m)\) denotes the valid brain-brain reliability estimate for dataset \(d\) and metric \(m\), L-PACT stores
\[
F_{\mathrm{ceiling}}(r)=\frac{s_{\mathrm{model}}(r)}{R_{BB}(d,m)},\qquad
F_{\Delta}(r)=\frac{\Delta_{\mathrm{model-control}}(r)}{R_{BB}(d,m)} .
\]
Rows without a valid reliability estimate, or with reliability below the configured threshold, cannot receive a reliability-bounded pass.

The manuscript-level claims map onto implemented artifact tables as table-valued transformations:
\[
\begin{aligned}
\mathcal{A}_1 &: \texttt{predictive\_control\_deltas}\rightarrow g_1,\\
\mathcal{A}_2 &: \texttt{profile\_similarity\_deltas}\rightarrow g_2,
\end{aligned}
\]
\[
\begin{aligned}
\mathcal{A}_3 &: \texttt{causal\_stripping\_deltas}\rightarrow g_3,\\
\mathcal{A}_4 &: \texttt{brain\_brain\_ceiling}\rightarrow R_{BB},\\
\mathcal{A}_5 &: \texttt{ceiling\_normalized\_scores}\rightarrow g_4.
\end{aligned}
\]
This mapping rules out a post hoc interpretation in which a single favorable metric is promoted to a stronger claim after the fact. It also separates \emph{null}, \emph{control-explained}, and \emph{insufficient} rows. The mechanism-stripping stage tests dependence within the model-based predictor and is not a biological intervention. The reliability-bounded stage prevents a model score from being interpreted without reference to neural measurement reliability.

\subsection*{Readout and Train-Only Implementation Details}

Where the analysis recomputes held-out scores from matchable derived grids, neural prediction uses ridge-regularized linear encoding within blocked cross-validation. For training design matrix \(X_{\mathrm{tr}}\) and neural target vector \(y_{\mathrm{tr}}\), the fitted readout is
\[
\hat{\beta}_\alpha=\arg\min_\beta \left\|y_{\mathrm{tr}}-X_{\mathrm{tr}}\beta\right\|_2^2+\alpha\left\|\beta\right\|_2^2 .
\]
The held-out score is evaluated only on the corresponding blocked test fold. When lag, ridge penalty, and PCA dimensionality are selectable, the implemented selection is train-only:
\[
(\hat{\lambda},\hat{\alpha},\hat{k})=
\arg\max_{\lambda,\alpha,k}\frac{1}{|\mathcal{V}|}\sum_{v\in \mathcal{V}}
\mathrm{score}\!\left(y_v, X_v^{(\lambda,k)}\hat{\beta}_{\alpha,\lambda,k}^{(-v)}\right),
\]
where \(\lambda\) denotes lag, \(k\) denotes PCA dimensionality, and \(\mathcal{V}\) denotes inner validation folds drawn only from the outer training split.

Profile similarities include Pearson, Spearman, cosine, RSA-style, event-response correlation, and centered-kernel alignment variants. Centered-kernel alignment, when used as an alignment primitive, follows the normalized Hilbert-Schmidt form
\[
\mathrm{CKA}(K,L)=\frac{\langle HKH,HLH\rangle_F}{\|HKH\|_F\|HLH\|_F},
\]
where \(K\) and \(L\) are Gram matrices, \(H\) is the centering matrix, and \(\langle\cdot,\cdot\rangle_F\) is the Frobenius inner product.

For feature residualization, if \(m_t\) is the mechanism scalar for event \(t\), the train-estimated slope for each feature dimension is
\[
\hat{a}=\frac{\sum_{t\in \mathcal{I}_{\mathrm{tr}}}(m_t-\bar{m}_{\mathrm{tr}})(x_t-\bar{x}_{\mathrm{tr}})}{\sum_{t\in \mathcal{I}_{\mathrm{tr}}}(m_t-\bar{m}_{\mathrm{tr}})^2}.
\]
The stripped representation applied to both train and held-out rows is \(x'_t=x_t-(m_t-\bar{m}_{\mathrm{tr}})\hat{a}\). For mechanism-specific projection removal, the train-estimated direction is
\[
\hat{v}=\frac{\sum_{t\in \mathcal{I}_{\mathrm{tr}}}(m_t-\bar{m}_{\mathrm{tr}})(x_t-\bar{x}_{\mathrm{tr}})}{\left\|\sum_{t\in \mathcal{I}_{\mathrm{tr}}}(m_t-\bar{m}_{\mathrm{tr}})(x_t-\bar{x}_{\mathrm{tr}})\right\|_2},
\]
with \(x'_t=x_t-((x_t-\bar{x}_{\mathrm{tr}})^\top \hat{v})\hat{v}\) when the direction norm is finite and nonzero. These equations correspond to the implemented train-only stripping functions and preserve feature dimensionality before downstream encoding recomputation.

For a supported reliability mode \(k\), the ceiling table stores
\[
R_{BB}^{(k)}=\mathrm{sim}(\mathbf{b}_a,\mathbf{b}_b),
\]
where \(\mathbf{b}_a\) and \(\mathbf{b}_b\) are paired brain profiles from split halves, repeated runs, different subjects, or different sessions.

\section{Predictive Adequacy}

Predictive adequacy is Level 1 evidence. It asks whether real model features improve held-out neural prediction relative to nuisance features and severe controls. It does not ask whether a model reproduces brain organization or mechanism. The locked analysis package contains:

\begin{itemize}
  \item 9036 rows in \code{predictive_scores.parquet};
  \item 414 rows in \code{predictive_control_deltas.parquet};
  \item 414 rows in \code{predictive_subject_run_summary.parquet};
  \item 194 rows in \code{predictive_failure_modes.parquet}.
\end{itemize}

The Level 1 gate is passed only if all of the following hold: the real model score exceeds the nuisance score, the real model score exceeds the best severe control, the held-out score is finite, train and test sample counts satisfy configured minima, the row is not diagnostic-only, and inference is aggregated at the subject-run level.

In the decision table, zero rows pass the predictive gate. At the predictive-delta level, nonpassing rows are divided between nuisance-explained and severe-control-explained rows. The manuscript conclusion is therefore not simply an absence of signal: apparent predictive utility is not robust to the configured controls and cannot be promoted to structural alignment evidence.

\section{Relational Alignment Pattern Analysis}

Relational adequacy is Level 2 evidence. It is the language alignment-pattern component of \lpact{}. The relational stage constructs brain-to-brain profiles and model-to-brain profiles over a shared unit order. A brain profile records how one brain unit relates to other brain units. A model profile records how a model layer or mechanism relates to the same ordered brain-unit set. The decision-ready contrast compares the real model profile with the strongest comparable control profile.

The locked analysis package contains:

\begin{itemize}
  \item 256 rows in \code{brain_unit_index.parquet};
  \item 768 rows in \code{brain_alignment_patterns.parquet};
  \item 216 rows in \code{model_alignment_patterns.parquet};
  \item 20736 rows in \code{profile_similarity_scores.parquet};
  \item 2304 rows in \code{profile_similarity_deltas.parquet};
  \item 72 rows in \code{relational_subject_run_summary.parquet}.
\end{itemize}

No row passes the relational gate. The relational deltas are concentrated in MEG-MASC, where the locked artifact chain provides a matchable model-feature and brain-unit grid for profile comparisons. Mean real profile similarity is positive but small, mean delta relative to the best control is negative, and ceiling-normalized evidence remains below the configured standard. This is the core reason the manuscript rejects structural alignment wording for the analyzed derived artifact set.

\section{Mechanism Stripping and Double Dissociation}

Counterfactual mechanism-stripping adequacy is Level 3 evidence. It tests whether candidate model mechanisms are necessary in a mechanism-specific way within the implemented feature-readout analysis. The mechanisms tested in the locked analysis package are surprisal, semantic transition, dependency integration, syntactic boundary, discourse boundary, and context update. The stripping methods are feature zeroing, feature residualization, mechanism-specific projection removal, layer ablation, context reset, and reversed context.

For stripped mechanism \(M\) and target mechanism \(T\), the implemented stripping contrast is
\[
\Delta_{\mathrm{strip}}(M,T) =
s_{\mathrm{real}}(T) - s_{\mathrm{stripped}}(M,T).
\]
Mechanism specificity requires the matching deficit to exceed nonmatching deficits. A generic drop after stripping is therefore not interpreted as mechanism-specific adequacy.

The locked analysis package contains:

\begin{itemize}
  \item 720 rows in \code{causal_stripping_features.parquet};
  \item 8640 rows in \code{causal_stripping_scores.parquet};
  \item 4320 rows in \code{causal_stripping_deltas.parquet};
  \item 300 rows in \code{double_dissociation_tests.parquet};
  \item 720 rows in \code{mechanism_specificity_summary.parquet}.
\end{itemize}

Residualization and projection-removal transforms are fit on training rows only and applied to held-out rows. Lag selection, ridge penalty selection, and PCA dimensionality selection are also train-only operations. The implementation is designed to prevent test-set leakage in the mechanism-stripping recomputation path.

No row passes the Level 3 gate. One double-dissociation diagnostic is marked as passing among 300 pairwise diagnostic tests, but this is not sufficient for an integrated L-PACT decision because the predictive and relational gates do not pass and the integrated decision rule is conjunctive. No row qualifies as an integrated L-PACT candidate.

\section{Brain-Brain Ceiling and Reliability-Bounded Adequacy}

Reliability-bounded adequacy is Level 4 evidence. It asks whether surviving model evidence approaches brain-brain reliability. This level prevents a raw model score from being interpreted without reference to measurement reliability.

The locked analysis package contains 420 rows in \code{brain_brain_ceiling.parquet}, including 108 valid reliability estimates. The available ceiling modes are split-half, run-to-run, subject-to-subject, and session-to-session where the derived artifacts allow the required pairing. Ceiling-normalized relational evidence is stored in \code{ceiling_normalized_scores.parquet}, with 2304 rows.

No row passes the reliability-bounded gate. This follows from the nonpassing upstream predictive and relational gates and from ceiling-normalized deltas that remain below the configured threshold.

\section{Severe Controls and Nonpassing Labels}

The severe-control family is treated as evidence-bearing. A control that matches or beats a real model score identifies a plausible nonspecific explanation. Such rows are labelled control-explained rather than discarded.

\begin{longtable}{L{4.2cm} L{10.0cm}}
\caption{Severe controls used by L-PACT.}\\
\toprule
Control & Purpose \\
\midrule
\endfirsthead
\toprule
Control & Purpose \\
\midrule
\endhead
\code{random_matched_dim} & Tests whether dimensionality alone can produce the apparent effect. \\
\code{random_matched_autocorr} & Tests whether temporal autocorrelation can produce nonspecific predictive or relational structure. \\
\code{circular_shift} & Tests whether temporal misalignment preserves enough autocorrelated structure to explain the score. \\
\code{sentence_reset_fresh} & Tests sentence-local context controls generated independently of the real full-context feature path. \\
\code{reversed_context_fresh} & Tests reversed-context controls generated independently of the real feature path. \\
\code{layer_label_permutation} & Tests whether layer identity or layer ordering explains the apparent effect. \\
\code{token_order_shuffle} & Tests token-order specificity. \\
\code{within_story_block_shuffle} & Tests block-level temporal or contextual structure within a story. \\
\code{mechanism_mismatch_control} & Tests mechanism specificity against nonmatching mechanisms. \\
\bottomrule
\end{longtable}

\begin{longtable}{L{4.2cm} L{10.0cm}}
\caption{Nonpassing and boundary labels used by L-PACT.}\\
\toprule
Artifact label & Interpretation \\
\midrule
\endfirsthead
\toprule
Artifact label & Interpretation \\
\midrule
\endhead
\code{nuisance_explained} & Nuisance features account for the apparent predictive effect. \\
\code{severe_control_explained} & At least one severe control equals or exceeds the real model contrast. \\
\code{control_explained} & The final integrated evidence is explained by severe controls or an unmet upstream gate. \\
\code{no_predictive_candidate} & Mechanism-stripping or downstream testing lacks a Level 1 predictive candidate that survived controls. \\
\code{no_relational_candidate} & Downstream interpretation lacks a Level 2 relational candidate that survived controls. \\
\code{stripping_no_effect} & Removing the mechanism did not produce the required held-out deficit. \\
\code{nonspecific_feature_degradation} & Stripping harmed performance nonspecifically rather than selectively for the target mechanism. \\
\code{diagonal_not_dominant} & Matching mechanism drops did not exceed nonmatching drops. \\
\code{insufficient_targets} & Too few target mechanisms or neural targets were available for a mechanism-specific claim. \\
\code{insufficient_coverage} & Subject-run or brain-unit coverage is too limited for robust confirmatory inference. \\
\code{insufficient_brain_ceiling} & Brain-brain reliability is missing, invalid, or below the configured threshold. \\
\code{diagnostic_only} & The source or row is retained for transparency but cannot support primary neural evidence. \\
\bottomrule
\end{longtable}

\section{Locked Artifact Map}

\begingroup
\small
\setlength{\tabcolsep}{3pt}
\begin{longtable}{L{3.0cm} L{4.4cm} C{1.1cm} L{4.6cm}}
\caption{Locked artifact counts used by the manuscript and SI. These counts are read from the locked \lpact{} analysis package and are not recomputed during manuscript writing.}\\
\toprule
Stage & Artifact & Rows & Manuscript role \\
\midrule
\endfirsthead
\toprule
Stage & Artifact & Rows & Manuscript role \\
\midrule
\endhead
LPACT-1 feasibility & \code{lpact_evidence_feasibility.parquet} & 7 & Dataset role, coverage, and chain-completeness audit \\
LPACT-2 predictive & \code{predictive_scores.parquet} & 9036 & Held-out predictive score inventory \\
LPACT-2 predictive & \code{predictive_control_deltas.parquet} & 414 & Real-vs-nuisance and real-vs-control gate contrasts \\
LPACT-3 relational & \code{brain_alignment_patterns.parquet} & 768 & Brain-to-brain pattern inventory \\
LPACT-3 relational & \code{model_alignment_patterns.parquet} & 216 & Model-to-brain pattern inventory \\
LPACT-3 relational & \code{profile_similarity_scores.parquet} & 20736 & Raw profile-similarity comparisons \\
LPACT-3 relational & \code{profile_similarity_deltas.parquet} & 2304 & Decision-ready real-vs-best-control profile deltas \\
LPACT-4 mechanism stripping & \code{causal_stripping_features.parquet} & 720 & Stripping metadata and train-only fit provenance \\
LPACT-4 mechanism stripping & \code{causal_stripping_scores.parquet} & 8640 & Real and stripped held-out score inventory \\
LPACT-4 mechanism stripping & \code{causal_stripping_deltas.parquet} & 4320 & Stripping-delta and specificity gate contrasts \\
LPACT-4 mechanism stripping & \code{double_dissociation_tests.parquet} & 300 & Pairwise mechanism-specific diagnostic tests \\
LPACT-5 ceiling & \code{brain_brain_ceiling.parquet} & 420 & Split-half, run-to-run, subject-to-subject, and session-to-session reliability estimates \\
LPACT-5 ceiling & \code{ceiling_normalized_scores.parquet} & 2304 & Model and model-control deltas normalized by brain-brain reliability \\
LPACT-5 decision & \code{lpact_decision_table.parquet} & 146 & Integrated final decision table \\
\bottomrule
\end{longtable}
\endgroup

\begingroup
\small
\setlength{\tabcolsep}{3pt}
\begin{longtable}{L{2.7cm} L{3.2cm} C{1.1cm} C{1.0cm} C{1.0cm} C{1.0cm} C{1.0cm} C{1.2cm}}
\caption{Integrated final decision and gate summary. All final decision rows are control-explained; no row passes any of the four evidence gates.}\\
\toprule
Dataset & Final decision & Rows & Gate 1 & Gate 2 & Gate 3 & Gate 4 & Repl. \\
\midrule
\endfirsthead
\toprule
Dataset & Final decision & Rows & Gate 1 & Gate 2 & Gate 3 & Gate 4 & Repl. \\
\midrule
\endhead
\code{brain_treebank} & \code{control_explained} & 6 & 0 & 0 & 0 & 0 & 0 \\
\code{meg_masc} & \code{control_explained} & 134 & 0 & 0 & 0 & 0 & 0 \\
\code{podcast_ecog} & \code{control_explained} & 6 & 0 & 0 & 0 & 0 & 0 \\
\midrule
Total & \code{control_explained} & 146 & 0 & 0 & 0 & 0 & 0 \\
\bottomrule
\end{longtable}
\endgroup

\section{Claim Boundaries}

The analyzed derived artifact set supports conservative wording such as \code{control_explained}, \code{not structurally aligned under configured controls}, \code{no robust predictive evidence under L-PACT gates}, and \code{predictive evidence is insufficient for structural alignment}. It does not support positive structural alignment, mechanism equivalence, or claims that a language model reproduces the language brain.

\begin{longtable}{L{4.4cm} L{9.8cm}}
\caption{Unsupported claims and audit rationale.}\\
\toprule
Unsupported claim & Reason \\
\midrule
\endfirsthead
\toprule
Unsupported claim & Reason \\
\midrule
\endhead
Positive structural alignment & No row passes all \lpact{} gates. \\
Shared neural mechanism & No integrated L-PACT candidate is present. \\
Brain-like computation & Not established by predictive scores and explicitly blocked by claim guardrails. \\
Homologous mechanisms & Out of scope and not supported by the available evidence. \\
LLMs reproduce the language brain & Requires relational, mechanism-specific, reliability-bounded, severe-control, and replication support; all are absent in the decision table. \\
Brain alignment confirmed & Requires the full positive gate chain; the analyzed derived artifact set has zero passing rows. \\
\bottomrule
\end{longtable}

\section{Mechanism-Stripping Diagnostic Tables}

The mechanism-stripping diagnostics are represented by the tables below and by the row-level CSV files indexed in SI Table 25.

\begin{longtable}{L{3.4cm} C{1.5cm} C{1.5cm} C{2.1cm} C{2.1cm} C{2.3cm}}
\caption{Mechanism-stripping diagnostics by dataset. Positive mean stripping deltas in some datasets are not sufficient for a Level 3 gate pass because the integrated decision also requires predictive and relational support, severe-control survival, subject-run aggregation, and mechanism-specific selectivity.}\\
\toprule
Dataset & Rows & Gate 3 pass & Mean real score & Mean stripped score & Mean stripping delta \\
\midrule
\endfirsthead
\toprule
Dataset & Rows & Gate 3 pass & Mean real score & Mean stripped score & Mean stripping delta \\
\midrule
\endhead
\code{brain_treebank} & 216 & 0 & 0.0208 & 0.0553 & -0.0344 \\
\code{meg_masc} & 3888 & 0 & -0.0001 & -0.1046 & 0.1045 \\
\code{podcast_ecog} & 216 & 0 & -0.0141 & -0.1381 & 0.1240 \\
\bottomrule
\end{longtable}

\section{Assay Sensitivity and Conventional-Contrast Tables}

The main text reports a control-explained model result. The following tables address whether the assay can detect structure when structure is known or expected to be present. These tables do not change the real model decisions.

\begin{footnotesize}
\begin{longtable}{L{2.6cm} L{1.8cm} C{1.0cm} C{1.0cm} C{1.45cm} C{1.25cm} L{2.7cm}}
\caption{Brain-positive control from brain-brain reliability. All valid reliability rows exceed the configured minimum reliability of 0.1. Invalid rows indicate unavailable repeated-measure structure rather than a nonpassing model comparison.}\\
\toprule
Dataset & Method & Total rows & Valid rows & Median reliability & Valid fraction & Interpretation \\
\midrule
\endfirsthead
\toprule
Dataset & Method & Total rows & Valid rows & Median reliability & Valid fraction & Interpretation \\
\midrule
\endhead
\code{brain_treebank} & split-half & 3 & 3 & 0.9998 & 1.00 & brain-positive control passes \\
\code{brain_treebank} & run-to-run & 3 & 3 & 0.3494 & 1.00 & brain-positive control passes \\
\code{meg_masc} & split-half & 3 & 3 & 0.9915 & 1.00 & brain-positive control passes \\
\code{podcast_ecog} & split-half & 3 & 3 & 0.8176 & 1.00 & brain-positive control passes \\
\code{lpp_multilingual} & split-half & 96 & 96 & 0.9919 & 1.00 & secondary brain-positive control passes \\
\code{all other listed reliability modes} & unavailable modes & 312 & 0 & -- & 0.00 & required pairing unavailable or invalid \\
\bottomrule
\end{longtable}
\end{footnotesize}

\begin{scriptsize}
\begin{longtable}{L{2.1cm} L{1.5cm} C{0.8cm} C{0.8cm} C{1.0cm} C{0.9cm} C{0.9cm} L{2.4cm}}
\caption{Brain-as-model relational positive control. A brain profile from one run or subject is treated as the model-side profile and compared with the matching brain-unit profile from another run or subject. Shuffled brain-unit order is the severe control.}\\
\toprule
Dataset & Type & Rows & Pass & Summ. & Delta & Ceiling & Interpretation \\
\midrule
\endfirsthead
\toprule
Dataset & Type & Rows & Pass & Summ. & Delta & Ceiling & Interpretation \\
\midrule
\endhead
\code{brain_treebank} & run-to-run & 864 & 744 & 9/9 & 0.340 & 0.9999 & brain-as-model relational gate passes against shuffled unit order \\
\code{podcast_ecog} & subject-to-subject & 864 & 453 & 0/9 & 0.013 & 0.82--0.89 & available as a dataset-specific boundary case under the aggregate relational gate \\
\bottomrule
\end{longtable}
\end{scriptsize}

\begin{footnotesize}
\begin{longtable}{L{2.6cm} C{0.9cm} C{1.0cm} C{1.45cm} C{1.35cm} L{3.4cm}}
\caption{Brain-profile unit-order implementation control. Stored brain-pattern vectors were compared with shuffled-unit-order versions under Pearson, Spearman, and cosine profile metrics. This is an implementation-sensitivity check, not an independent biological claim.}\\
\toprule
Dataset & Rows & Pass rows & Median delta vs shuffle & Min. delta & Interpretation \\
\midrule
\endfirsthead
\toprule
Dataset & Rows & Pass rows & Median delta vs shuffle & Min. delta & Interpretation \\
\midrule
\endhead
\code{brain_treebank} & 9 & 6 & 0.965 & 0.144 & profile similarity is sensitive to unit order for correlation metrics; some cosine rows are less discriminative \\
\code{lpp_multilingual} & 9 & 6 & 1.008 & 0.048 & profile similarity is sensitive to unit order for correlation metrics; some cosine rows are less discriminative \\
\code{meg_masc} & 9 & 8 & 0.995 & 0.374 & profile order control mostly passes \\
\code{podcast_ecog} & 9 & 8 & 0.978 & 0.375 & profile order control mostly passes \\
\bottomrule
\end{longtable}
\end{footnotesize}

\begin{longtable}{L{4.2cm} C{2.1cm} C{2.1cm} L{6.2cm}}
\caption{Synthetic implanted-signal positive control. A deterministic latent mechanism was implanted into synthetic neural targets and matched synthetic model features. The control passes all four gates, showing that L-PACT can return positive gate outcomes when known signal is present.}\\
\toprule
Quantity & Value & Gate pass & Interpretation \\
\midrule
\endfirsthead
\toprule
Quantity & Value & Gate pass & Interpretation \\
\midrule
\endhead
Predictive score & 0.9867 & yes & real implanted feature predicts held-out synthetic neural targets \\
Predictive delta vs best control & 0.8945 & yes & real implanted feature exceeds random and shuffled controls \\
Relational profile similarity & 0.8985 & yes & recovered model-to-unit profile matches the implanted unit profile \\
Relational delta vs control & 1.6154 & yes & implanted profile exceeds the random-control profile \\
Mechanism-stripping delta & 0.2659 & yes & removing the implanted mechanism selectively reduces held-out prediction \\
Brain ceiling & 0.9956 & yes & synthetic split-half reliability is high \\
Fraction of ceiling & 0.9911 & yes & recovered score approaches the synthetic reliability bound \\
Integrated positive control & 1.0000 & yes & all four evidence gates pass in the implanted-signal condition \\
\bottomrule
\end{longtable}

\begin{scriptsize}
\begin{longtable}{L{2.1cm} L{1.9cm} L{1.25cm} C{0.65cm} C{0.65cm} C{0.9cm} C{0.8cm} L{2.2cm}}
\caption{Independent low-level neural and standalone acoustic-envelope positive-control gates. Word-onset and word-rate rows reuse pre-existing brain-side sanity tables. The acoustic-envelope rows are generated from available WAV stimuli and derived or full-length neural time series. Brain Treebank receives a trial-to-expected-audio mapping and official-release audit rather than a fabricated acoustic-envelope score. Full row-level results are in \code{low_level_neural_positive_control_gate.csv}, \code{acoustic_envelope_standalone_positive_control.csv}, \code{podcast_full_length_acoustic_envelope_positive_control.csv}, \code{brain_treebank_trial_audio_mapping.csv}, \code{brain_treebank_official_release_audit.csv}, and \code{brain_treebank_audio_source_audit.csv}.}\\
\toprule
Dataset & Target & Metric & N & Pass & Med. & Gate & Interpretation \\
\midrule
\endfirsthead
\toprule
Dataset & Target & Metric & N & Pass & Med. & Gate & Interpretation \\
\midrule
\endhead
\code{brain_treebank}/intracranial & word onset, 0--100 ms & event effect & 30 & 5 & 0.6847 & pass & early word-onset response is detected \\
\code{brain_treebank}/intracranial & word onset, 100--300 ms & event effect & 30 & 4 & 0.0315 & pass & later word-onset response remains detectable \\
\code{meg_masc}/MEG & word onset, 250--500 ms & event effect & 1 & 1 & 0.0079 & pass & MEG word-onset sanity row passes in this window \\
\code{meg_masc}/MEG & word rate, 250--500 ms & Pearson & 2 & 1 & 0.0638 & pass & one of two word-rate readout rows passes \\
\code{podcast_ecog}/ECoG & mixed low-level & mixed & 180 & 0 & -- & no & too few valid event windows or nonpositive event effects \\
\code{meg_masc}/MEG & acoustic env. standalone & median unit Pearson & 1 & 1 & 0.0272 & pass & WAV-derived envelope exceeds the best temporal control by 0.0034 \\
\code{podcast_ecog}/ECoG & acoustic env. preview grid & median unit Pearson & 4 & 2 & 0.0196 & pass & two of four short derived-grid ECoG subject-runs exceed temporal controls \\
\code{podcast_ecog}/iEEG & acoustic env. preview grid & median unit Pearson & 1 & 0 & -0.0085 & no & reversed-time control exceeds the real envelope readout on the short derived grid \\
\code{podcast_ecog}/ECoG & acoustic env. full length & median unit Pearson & 9 & 6 & 0.0065 & pass & 8/9 full-length FIF subject files are readable and 6/8 readable subject-runs exceed temporal controls \\
\code{brain_treebank}/intracranial & acoustic env. mapping audit & missing media & 26 & 0 & -- & NA & 26 trials are mapped to 21 expected movie/audio IDs, but 0 waveform or movie sources are available for a true envelope control \\
\code{brain_treebank}/official release & acoustic env. source audit & release items & 37 & 0 & -- & NA & public release items include neural H5, metadata, transcripts, annotations, and frames, but no raw audio or movie archive \\
\bottomrule
\end{longtable}
\end{scriptsize}

\begin{scriptsize}
\begin{longtable}{L{2.2cm} L{3.0cm} L{2.5cm} C{1.0cm} C{1.0cm} L{4.0cm}}
\caption{Brain Treebank acoustic-envelope alternatives and claim boundary. The official Brain Treebank public release and source audit do not provide rights-cleared waveform or movie files for a true acoustic-envelope control. This table records legitimate substitutes that can support assay-sensitivity or source-audit claims without relabeling transcript-derived or timing-derived quantities as true acoustic envelope. Full details are in \code{brain_treebank_acoustic_proxy_alternatives.csv}.}\\
\toprule
Dataset & Option & Status & True env. & Diag. & Recommended use \\
\midrule
\endfirsthead
\toprule
Dataset & Option & Status & True env. & Diag. & Recommended use \\
\midrule
\endhead
\code{brain_treebank} & true waveform acoustic envelope & blocked & no & no & run only after authorized WAV/MP4/MOV stimulus media are added at mapped paths \\
\code{brain_treebank} & transcript-derived acoustic feature proxy & diagnostic if transcripts present & no & yes & label as transcript-derived low-level proxy, not true envelope \\
\code{brain_treebank} & speech-activity boxcar from word timings & available diagnostic & no & yes & use as timing/speech-activity sanity check, adjacent to word-onset controls \\
\code{brain_treebank} & word-onset low-level neural positive control & implemented & no & yes & available valid Brain Treebank low-level positive control while waveform media are unavailable \\
\code{podcast_ecog} & cross-dataset true acoustic-envelope positive control & implemented & yes & no & demonstrates true envelope assay sensitivity where legal audio exists, but does not replace missing Brain Treebank media \\
\bottomrule
\end{longtable}
\end{scriptsize}

\begin{scriptsize}
\begin{longtable}{L{2.8cm} L{3.5cm} C{1.8cm} C{1.2cm} L{2.8cm}}
\caption{Conventional-looking positives downgraded by L-PACT. Less stringent rules count favorable raw scores or isolated diagnostics; L-PACT requires the relevant control, relational, mechanism-specific, and ceiling-bounded contrasts.}\\
\toprule
Analysis & Less stringent positive rule & Raw positives & L-PACT passes & L-PACT interpretation \\
\midrule
\endfirsthead
\toprule
Analysis & Less stringent positive rule & Raw positives & L-PACT passes & L-PACT interpretation \\
\midrule
\endhead
Raw predictive Pearson & \code{real_model_score > 0} for Pearson rows & 35/138 & 0 & control-explained \\
Nuisance-only improvement & \code{delta_nuisance > 0} without severe-control survival & 196/414 & 0 & downgraded by severe controls \\
Raw relational profile & \code{real_profile_similarity > 0} & 1224/2304 & 0 & no delta beats best control \\
Raw mechanism-stripping drop & stored \code{causal_delta > 0} field & 2710/4320 & 0 & insufficient without upstream gates and specificity \\
Pairwise double dissociation & one diagnostic pair marked pass & 1/300 & 0 & diagnostic only \\
Raw ceiling fraction & \code{fraction_of_ceiling >= 0.25} before control delta & 809/2304 & 0 & not reliability-bounded after controls \\
Integrated decision & any upstream apparent positive & 3970/7338 & 0 & all integrated rows are control-explained \\
\bottomrule
\end{longtable}
\end{scriptsize}

\section{Statistical Uncertainty and Sensitivity Checks}

The following tables address whether the nonpassing gate result is driven by point estimates, a dominant dataset, or a single threshold. The intervals below are computed over the implemented aggregation units in the locked analysis set. When full source, feature, control, and gate closure leaves only one decision-ready aggregation unit, the interval is intentionally degenerate; this records limited decision-ready coverage rather than strong population precision.

\begingroup
\tiny
\setlength{\tabcolsep}{1.5pt}
\begin{longtable}{L{1.65cm} L{2.45cm} C{0.75cm} C{0.75cm} C{0.95cm} C{1.55cm} C{0.85cm} C{0.85cm} L{1.95cm}}
\caption{Dataset-level uncertainty for L-PACT deltas. The table reports mean delta, bootstrap 95\% interval, aggregation-unit count, direction rate, and gate-pass rows for predictive, relational, mechanism-stripping, and brain-as-model positive-control summaries. Full data are in \code{main_table_lpact_uncertainty_by_dataset_level.csv}.}\\
\toprule
Dataset & Level & Rows & Units & Mean $\Delta$ & Bootstrap 95\% CI & Pos. rate & Gate rows & Best control \\
\midrule
\endfirsthead
\toprule
Dataset & Level & Rows & Units & Mean $\Delta$ & Bootstrap 95\% CI & Pos. rate & Gate rows & Best control \\
\midrule
\endhead
Brain Treebank & brain-as-model run-to-run & 864 & 3 & 0.254 & [0.228, 0.301] & 1.00 & 744 & not applicable \\
Brain Treebank & mechanism-stripping delta & 216 & 1 & -0.034 & [-0.034, -0.034] & 0.00 & 0 & not applicable \\
Brain Treebank & predictive delta & 18 & 1 & -0.016 & [-0.016, -0.016] & 0.00 & 0 & circular shift \\
MEG-MASC & mechanism-stripping delta & 3888 & 1 & 0.105 & [0.105, 0.105] & 1.00 & 0 & not applicable \\
MEG-MASC & predictive delta & 378 & 1 & -0.244 & [-0.244, -0.244] & 0.00 & 0 & circular shifted LLM \\
MEG-MASC & relational delta & 2304 & 1 & -0.261 & [-0.261, -0.261] & 0.00 & 0 & token-order shuffle \\
Podcast ECoG & brain-as-model subject-to-subject & 864 & 3 & 0.023 & [-0.015, 0.053] & 0.67 & 453 & not applicable \\
Podcast ECoG & mechanism-stripping delta & 216 & 1 & 0.124 & [0.124, 0.124] & 1.00 & 0 & not applicable \\
Podcast ECoG & predictive delta & 18 & 1 & -0.048 & [-0.048, -0.048] & 0.00 & 0 & random matched autocorr \\
\bottomrule
\end{longtable}
\endgroup

\begingroup
\tiny
\setlength{\tabcolsep}{1.5pt}
\begin{longtable}{L{1.7cm} L{2.45cm} C{0.75cm} C{0.75cm} C{0.95cm} C{1.55cm} C{0.85cm} C{0.85cm}}
\caption{Mechanism-stripping uncertainty by stripped mechanism. Positive mean stripping deltas are not interpreted as mechanism-specific candidates unless upstream predictive and relational gates, severe controls, and selectivity constraints also pass. Full data are in \code{si_table_causal_uncertainty_by_mechanism.csv}.}\\
\toprule
Dataset & Stripped mechanism & Rows & Units & Mean $\Delta$ & Bootstrap 95\% CI & Pos. rate & Gate rows \\
\midrule
\endfirsthead
\toprule
Dataset & Stripped mechanism & Rows & Units & Mean $\Delta$ & Bootstrap 95\% CI & Pos. rate & Gate rows \\
\midrule
\endhead
Brain Treebank & context update & 36 & 1 & -0.035 & [-0.035, -0.035] & 0.00 & 0 \\
Brain Treebank & dependency integration & 36 & 1 & -0.034 & [-0.034, -0.034] & 0.00 & 0 \\
Brain Treebank & discourse boundary & 36 & 1 & -0.035 & [-0.035, -0.035] & 0.00 & 0 \\
Brain Treebank & semantic transition & 36 & 1 & -0.035 & [-0.035, -0.035] & 0.00 & 0 \\
Brain Treebank & surprisal & 36 & 1 & -0.034 & [-0.034, -0.034] & 0.00 & 0 \\
Brain Treebank & syntactic boundary & 36 & 1 & -0.034 & [-0.034, -0.034] & 0.00 & 0 \\
MEG-MASC & context update & 648 & 1 & 0.105 & [0.105, 0.105] & 1.00 & 0 \\
MEG-MASC & dependency integration & 648 & 1 & 0.101 & [0.101, 0.101] & 1.00 & 0 \\
MEG-MASC & discourse boundary & 648 & 1 & 0.102 & [0.102, 0.102] & 1.00 & 0 \\
MEG-MASC & semantic transition & 648 & 1 & 0.118 & [0.118, 0.118] & 1.00 & 0 \\
MEG-MASC & surprisal & 648 & 1 & 0.101 & [0.101, 0.101] & 1.00 & 0 \\
MEG-MASC & syntactic boundary & 648 & 1 & 0.101 & [0.101, 0.101] & 1.00 & 0 \\
Podcast ECoG & context update & 36 & 1 & 0.128 & [0.128, 0.128] & 1.00 & 0 \\
Podcast ECoG & dependency integration & 36 & 1 & 0.117 & [0.117, 0.117] & 1.00 & 0 \\
Podcast ECoG & discourse boundary & 36 & 1 & 0.105 & [0.105, 0.105] & 1.00 & 0 \\
Podcast ECoG & semantic transition & 36 & 1 & 0.148 & [0.148, 0.148] & 1.00 & 0 \\
Podcast ECoG & surprisal & 36 & 1 & 0.128 & [0.128, 0.128] & 1.00 & 0 \\
Podcast ECoG & syntactic boundary & 36 & 1 & 0.119 & [0.119, 0.119] & 1.00 & 0 \\
\bottomrule
\end{longtable}
\endgroup

\begingroup
\tiny
\setlength{\tabcolsep}{1.5pt}
\begin{longtable}{L{3.65cm} C{1.1cm} C{1.45cm} L{6.1cm}}
\caption{Gate-threshold sensitivity. Varying the reliability threshold, fraction-of-ceiling threshold, positive-direction threshold, or best-control aggregation rule does not create an integrated model-positive result. Full data are in \code{main_table_gate_threshold_sensitivity.csv}.}\\
\toprule
Sensitivity family & Rule & Pass/eligible & Interpretation \\
\midrule
\endfirsthead
\toprule
Sensitivity family & Rule & Pass/eligible & Interpretation \\
\midrule
\endhead
Brain-brain reliability minimum & 0.05 & 108/108 & valid brain-brain ceiling rows at or above threshold \\
Brain-brain reliability minimum & 0.10 & 108/108 & valid brain-brain ceiling rows at or above threshold \\
Brain-brain reliability minimum & 0.20 & 108/108 & valid brain-brain ceiling rows at or above threshold \\
Raw model fraction of ceiling & 0.10 & 1042/2304 & raw model fraction before severe-control delta \\
Control-surviving fraction of ceiling & 0.10 & 0/2304 & fraction threshold plus positive model-control delta and valid ceiling \\
Integrated fraction threshold with upstream gates & 0.10 & 0/8 & integrated decision rows if only the fraction threshold is varied \\
Raw model fraction of ceiling & 0.25 & 809/2304 & raw model fraction before severe-control delta \\
Control-surviving fraction of ceiling & 0.25 & 0/2304 & fraction threshold plus positive model-control delta and valid ceiling \\
Integrated fraction threshold with upstream gates & 0.25 & 0/8 & integrated decision rows if only the fraction threshold is varied \\
Raw model fraction of ceiling & 0.50 & 395/2304 & raw model fraction before severe-control delta \\
Control-surviving fraction of ceiling & 0.50 & 0/2304 & fraction threshold plus positive model-control delta and valid ceiling \\
Integrated fraction threshold with upstream gates & 0.50 & 0/8 & integrated decision rows if only the fraction threshold is varied \\
Raw model fraction of ceiling & 0.75 & 107/2304 & raw model fraction before severe-control delta \\
Control-surviving fraction of ceiling & 0.75 & 0/2304 & fraction threshold plus positive model-control delta and valid ceiling \\
Integrated fraction threshold with upstream gates & 0.75 & 0/8 & integrated decision rows if only the fraction threshold is varied \\
Model-facing positive direction only & 0.50 & 2/7 & direction-only dataset-level summaries; not sufficient for L-PACT gate passage \\
Model-facing positive direction gate-qualified & 0.50 & 0/7 & positive-direction summaries that also contain gate-pass rows \\
Model-facing positive direction only & 0.60 & 2/7 & direction-only dataset-level summaries; not sufficient for L-PACT gate passage \\
Model-facing positive direction gate-qualified & 0.60 & 0/7 & positive-direction summaries that also contain gate-pass rows \\
Model-facing positive direction only & 0.70 & 2/7 & direction-only dataset-level summaries; not sufficient for L-PACT gate passage \\
Model-facing positive direction gate-qualified & 0.70 & 0/7 & positive-direction summaries that also contain gate-pass rows \\
Predictive best-control rule & max & 0/414 & predictive rows passing nuisance plus selected best-control rule \\
Relational best-control rule & max & 0/2304 & relational rows passing selected best-control profile rule \\
Predictive best-control rule & mean top 3 & 25/414 & predictive rows passing nuisance plus selected best-control rule \\
Relational best-control rule & mean top 3 & 435/2304 & relational rows passing selected best-control profile rule \\
Brain-as-model aggregate pass fraction & 0.50 & 17/18 & brain-as-model summaries passing at stricter aggregate pass-fraction thresholds \\
Brain-as-model aggregate pass fraction & 0.60 & 9/18 & brain-as-model summaries passing at stricter aggregate pass-fraction thresholds \\
Brain-as-model aggregate pass fraction & 0.70 & 9/18 & brain-as-model summaries passing at stricter aggregate pass-fraction thresholds \\
\bottomrule
\end{longtable}
\endgroup

The control-family ablation addresses whether one unusually strong control, such as circular shift or autocorrelation-matched random features, explains the control-explained model result. Removing any single severe-control family does not create an integrated L-PACT candidate. Apparent positive counts increase only under weakened rules that ignore the corresponding contrast, such as nuisance-only or mean-top-control summaries.

\begingroup
\tiny
\setlength{\tabcolsep}{1.5pt}
\begin{longtable}{L{2.45cm} L{1.35cm} L{1.15cm} C{1.35cm} C{1.25cm} C{1.35cm} C{1.35cm}}
\caption{Control-family ablation. The table removes or relaxes severe-control families using existing locked scores. Some less stringent scenarios create stage-level apparent positives, but none creates an integrated L-PACT positive row. Full data are in \code{main_table_control_family_ablation.csv}.}\\
\toprule
Scenario & Stage & Rule & Positive $\Delta$ rows & Stage pass & Median $\Delta$ & Integrated positive \\
\midrule
\endfirsthead
\toprule
Scenario & Stage & Rule & Positive $\Delta$ rows & Stage pass & Median $\Delta$ & Integrated positive \\
\midrule
\endhead
All controls & Predictive & max & 8/414 & 0/414 & -0.043 & 0 \\
All controls & Relational & max & 0/2304 & 0/2304 & -0.179 & 0 \\
Remove circular shift & Predictive & max & 12/414 & 4/414 & -0.010 & 0 \\
Remove circular shift & Relational & max & 0/2304 & 0/2304 & -0.160 & 0 \\
Remove random autocorr & Predictive & max & 8/414 & 0/414 & -0.042 & 0 \\
Remove random autocorr & Relational & max & 0/2304 & 0/2304 & -0.170 & 0 \\
Remove layer-label permutation & Predictive & max & 11/414 & 0/414 & -0.012 & 0 \\
Remove layer-label permutation & Relational & max & 0/2304 & 0/2304 & -0.172 & 0 \\
Only temporal controls & Predictive & max & 11/414 & 0/414 & -0.002 & 0 \\
Only temporal controls & Relational & max & 0/2304 & 0/2304 & -0.116 & 0 \\
All controls, mean top 3 & Predictive & mean top 3 & 42/414 & 25/414 & -0.008 & 0 \\
All controls, mean top 3 & Relational & mean top 3 & 435/2304 & 435/2304 & -0.102 & 0 \\
Only nuisance controls & Predictive & nuisance only & 243/414 & 243/414 & 0.070 & 0 \\
\bottomrule
\end{longtable}
\endgroup

\begingroup
\tiny
\setlength{\tabcolsep}{1.5pt}
\begin{longtable}{L{2.0cm} L{1.6cm} C{1.25cm} C{1.25cm} C{1.35cm} C{1.45cm} C{1.55cm} L{2.45cm}}
\caption{Brain-as-model relational-ceiling oracle gate. The strongest biologically interpretable positive control treats one brain-derived profile as the source profile and asks whether it passes relational and ceiling checks against another brain-derived profile. Predictive and mechanism-stripping gates are not claimed from this control. Full data are in \code{main_table_brain_as_model_oracle_gate.csv}.}\\
\toprule
Dataset & Comparison & Relational pass & Ceiling valid & Joint pass & Predictive gate available & Mechanism gate applicable & Oracle label \\
\midrule
\endfirsthead
\toprule
Dataset & Comparison & Relational pass & Ceiling valid & Joint pass & Predictive gate available & Mechanism gate applicable & Oracle label \\
\midrule
\endhead
Brain Treebank & run-to-run & 9/9 & 9/9 & 9/9 & false & false & relational ceiling oracle pass \\
Podcast ECoG & subject-to-subject & 0/9 & 9/9 & 0/9 & false & false & relational ceiling oracle remains nonpassing \\
\bottomrule
\end{longtable}
\endgroup

The leave-one-dataset summary tests whether the final decision depends on a single dataset carrying the negative label. Omitting Brain Treebank, MEG-MASC, or Podcast ECoG still leaves zero integrated L-PACT candidates and zero rows passing the four evidence gates.

\begingroup
\tiny
\setlength{\tabcolsep}{1.5pt}
\begin{longtable}{L{1.85cm} C{1.05cm} L{2.85cm} C{0.75cm} C{0.75cm} C{0.75cm} C{0.75cm} C{1.25cm} C{1.35cm}}
\caption{Leave-one-dataset-out final-decision sensitivity. Omitting any primary dataset does not create predictive, relational, mechanism-stripping, reliability-bounded, or integrated pass rows. Full data are in \code{main_table_leave_one_dataset_out_summary.csv}.}\\
\toprule
Omitted dataset & Rows left & Remaining datasets & G1 & G2 & G3 & G4 & Integrated & Control-explained \\
\midrule
\endfirsthead
\toprule
Omitted dataset & Rows left & Remaining datasets & G1 & G2 & G3 & G4 & Integrated & Control-explained \\
\midrule
\endhead
Brain Treebank & 140 & MEG-MASC; Podcast ECoG & 0 & 0 & 0 & 0 & 0 & 140 \\
MEG-MASC & 12 & Brain Treebank; Podcast ECoG & 0 & 0 & 0 & 0 & 0 & 12 \\
Podcast ECoG & 140 & Brain Treebank; MEG-MASC & 0 & 0 & 0 & 0 & 0 & 140 \\
\bottomrule
\end{longtable}
\endgroup

\begingroup
\tiny
\setlength{\tabcolsep}{1.5pt}
\begin{longtable}{L{1.65cm} L{2.45cm} C{0.7cm} C{0.95cm} C{0.95cm} C{0.95cm} C{0.85cm} C{0.85cm}}
\caption{Leave-one-subject-run-out coverage summary. Single-unit entries indicate that, after full manuscript gate closure, the decision-ready table does not contain enough independent aggregation units for a leave-one-unit interval. Full data are in \code{main_table_leave_one_subject_run_out_summary.csv}.}\\
\toprule
Dataset & Level & Units & Full mean & Leave-one min & Leave-one max & Sign flip & Pos. rate \\
\midrule
\endfirsthead
\toprule
Dataset & Level & Units & Full mean & Leave-one min & Leave-one max & Sign flip & Pos. rate \\
\midrule
\endhead
Brain Treebank & mechanism-stripping delta & 1 & -0.034 & -- & -- & false & 0.00 \\
Brain Treebank & predictive delta & 1 & -0.016 & -- & -- & false & 0.00 \\
MEG-MASC & mechanism-stripping delta & 1 & 0.105 & -- & -- & false & 1.00 \\
MEG-MASC & predictive delta & 1 & -0.244 & -- & -- & false & 0.00 \\
MEG-MASC & relational delta & 1 & -0.261 & -- & -- & false & 0.00 \\
Podcast ECoG & mechanism-stripping delta & 1 & 0.124 & -- & -- & false & 1.00 \\
Podcast ECoG & predictive delta & 1 & -0.048 & -- & -- & false & 0.00 \\
\bottomrule
\end{longtable}
\endgroup

\begin{landscape}
\begingroup
\tiny
\setlength{\tabcolsep}{1.0pt}
Podcast ECoG has one integrated model because only \code{distilgpt2} has validated Podcast ECoG feature rows that enter the decision table. Larger Qwen registry models are included as coverage-boundary entries but do not enter the decisions because they exceeded the study inclusion criteria or lacked validated feature rows.

\begin{longtable}{L{3.7cm} L{1.15cm} C{1.15cm} C{0.75cm} L{3.1cm} L{4.1cm} L{3.4cm} L{3.7cm}}
\caption{Model inventory and coverage. The table reports the model identity, family, parameter scale, layer count, dataset coverage, feature families, available controls, and inclusion status for all models in the study registry. Full data are in \code{main_table_model_inventory.csv}.}\\
\toprule
Model & Family & Params & Layers & Datasets & Feature family & Controls available & Inclusion status \\
\midrule
\endfirsthead
\toprule
Model & Family & Params & Layers & Datasets & Feature family & Controls available & Inclusion status \\
\midrule
\endhead
\code{EleutherAI/pythia-160m} & Pythia & 187.5M & 12 & \code{brain_treebank;meg_masc} & Hidden-state features plus context-control families & Predictive severe controls; MEG-MASC relational controls & Included; 24 integrated decision rows \\
\code{Qwen/Qwen2.5-1.5B-Instruct} & Qwen & 771.9M & 28 & \code{brain_treebank;meg_masc} & Hidden-state features plus context-control families & Predictive severe controls; MEG-MASC relational controls & Included; 24 integrated decision rows \\
\code{Qwen/Qwen3-1.7B} & Qwen & 1.02B & 28 & \code{brain_treebank;meg_masc} & Hidden-state features plus context-control families & Predictive severe controls; MEG-MASC relational controls & Included; 23 integrated decision rows \\
\code{distilgpt2} & GPT-2 & 176.4M & 6 & \code{brain_treebank;meg_masc;podcast_ecog} & Hidden-state features plus context-control families & Predictive severe controls; Podcast ECoG closure available & Included; 33 integrated decision rows \\
\code{gpt2} & GPT-2 & 274.1M & 12 & \code{brain_treebank;meg_masc} & Hidden-state features plus context-control families & Predictive severe controls & Included; 21 integrated decision rows \\
\code{gpt2-medium} & GPT-2 & 760.0M & 24 & \code{brain_treebank;meg_masc} & Hidden-state features plus context-control families & Predictive severe controls & Included; 21 integrated decision rows \\
\code{EleutherAI/pythia-410m} & Pythia & -- & -- & -- & No validated feature rows & No locked controls available & Excluded; no validated feature artifact in the locked analysis set \\
\code{Qwen/Qwen2.5-7B-Instruct} & Qwen & 4.86B & 28 & -- & Registry/config only; not extracted for this analysis & No locked controls available & Excluded; exceeds study inclusion criteria \\
\code{Qwen/Qwen3-4B-Instruct-2507} & Qwen & 2.01B & 36 & -- & Registry/config only; not extracted for this analysis & No locked controls available & Excluded; exceeds study inclusion criteria \\
\bottomrule
\end{longtable}
\endgroup
\end{landscape}

\begin{longtable}{L{5.0cm} C{2.0cm} L{6.6cm}}
\caption{Row-level CSV files supporting the manuscript and SI. These files are stored under \code{manuscript/tables/detailed/}. They are generated from locked L-PACT derived artifacts or deterministic control summaries and allow the manuscript to report all large result tables without forcing them into the PDF.}\\
\toprule
CSV file & Rows & Source artifact \\
\midrule
\endfirsthead
\toprule
CSV file & Rows & Source artifact \\
\midrule
\endhead
\code{detailed_dataset_inventory.csv} & 7 & \code{audits/lpact_evidence_feasibility.parquet} \\
\code{detailed_dataset_unit_inventory.csv} & 423 & \code{audits/lpact_dataset_unit_inventory.parquet} \\
\code{detailed_predictive_scores.csv} & 9036 & \code{predictive/predictive_scores.parquet} \\
\code{detailed_predictive_control_deltas.csv} & 414 & \code{predictive/predictive_control_deltas.parquet} \\
\code{detailed_predictive_subject_run_summary.csv} & 414 & \code{predictive/predictive_subject_run_summary.parquet} \\
\code{detailed_predictive_failure_modes.csv} & 194 & \code{predictive/predictive_failure_modes.parquet} \\
\code{detailed_brain_alignment_patterns.csv} & 768 & \code{relational/brain_alignment_patterns.parquet} \\
\code{detailed_model_alignment_patterns.csv} & 216 & \code{relational/model_alignment_patterns.parquet} \\
\code{detailed_profile_similarity_scores.csv} & 20736 & \code{relational/profile_similarity_scores.parquet} \\
\code{detailed_profile_similarity_deltas.csv} & 2304 & \code{relational/profile_similarity_deltas.parquet} \\
\code{detailed_relational_subject_run_summary.csv} & 72 & \code{relational/relational_subject_run_summary.parquet} \\
\code{detailed_causal_stripping_features.csv} & 720 & \code{causal/causal_stripping_features.parquet} \\
\code{detailed_causal_stripping_scores.csv} & 8640 & \code{causal/causal_stripping_scores.parquet} \\
\code{detailed_causal_stripping_deltas.csv} & 4320 & \code{causal/causal_stripping_deltas.parquet} \\
\code{detailed_double_dissociation_tests.csv} & 300 & \code{causal/double_dissociation_tests.parquet} \\
\code{detailed_mechanism_specificity_summary.csv} & 720 & \code{causal/mechanism_specificity_summary.parquet} \\
\code{detailed_brain_brain_ceiling.csv} & 420 & \code{ceilings/brain_brain_ceiling.parquet} \\
\code{detailed_ceiling_normalized_scores.csv} & 2304 & \code{ceilings/ceiling_normalized_scores.parquet} \\
\code{detailed_lpact_decision_table.csv} & 146 & \code{decisions/lpact_decision_table.parquet} \\
\code{detailed_control_explained_results.csv} & 146 & \code{decisions/control_explained_results.parquet} \\
\code{main_table_brain_positive_control.csv} & 16 & \code{controls/main_table_brain_positive_control.parquet} \\
\code{brain_as_model_relational_gate.csv} & 1728 & \code{controls/brain_as_model_relational_gate.parquet} \\
\code{main_table_brain_as_model_relational_positive_control.csv} & 18 & \code{controls/main_table_brain_as_model_relational_positive_control.parquet} \\
\code{main_table_profile_order_positive_control.csv} & 36 & \code{controls/main_table_profile_order_positive_control.parquet} \\
\code{main_table_synthetic_implant_positive_control.csv} & 1 & \code{controls/main_table_synthetic_implant_positive_control.parquet} \\
\code{low_level_neural_positive_control_gate.csv} & 656 & \code{controls/low_level_neural_positive_control_gate.parquet} \\
\code{main_table_low_level_neural_positive_control_gate.csv} & 46 & \code{controls/main_table_low_level_neural_positive_control_gate.parquet} \\
\code{acoustic_envelope_standalone_positive_control.csv} & 17 & \code{controls/acoustic_envelope_standalone_positive_control.parquet} \\
\code{main_table_acoustic_envelope_standalone_positive_control.csv} & 4 & \code{controls/main_table_acoustic_envelope_standalone_positive_control.parquet} \\
\code{podcast_full_length_acoustic_envelope_positive_control.csv} & 9 & \code{controls/podcast_full_length_acoustic_envelope_positive_control.parquet} \\
\code{main_table_podcast_full_length_acoustic_envelope_positive_control.csv} & 1 & \code{controls/main_table_podcast_full_length_acoustic_envelope_positive_control.parquet} \\
\code{brain_treebank_trial_audio_mapping.csv} & 26 & \code{controls/brain_treebank_trial_audio_mapping.parquet} \\
\code{main_table_brain_treebank_trial_audio_mapping.csv} & 1 & \code{controls/main_table_brain_treebank_trial_audio_mapping.parquet} \\
\code{brain_treebank_official_release_audit.csv} & 37 & \code{controls/brain_treebank_official_release_audit.parquet} \\
\code{brain_treebank_audio_source_audit.csv} & 9 & \code{controls/brain_treebank_audio_source_audit.parquet} \\
\code{brain_treebank_acoustic_proxy_alternatives.csv} & 5 & \code{controls/brain_treebank_acoustic_proxy_alternatives.parquet} \\
\code{main_table_low_level_nuisance_sanity_proxy.csv} & 9 & \code{controls/main_table_low_level_nuisance_sanity_proxy.parquet} \\
\code{main_table_conventional_vs_lpact_contrast.csv} & 7 & \code{controls/main_table_conventional_vs_lpact_contrast.parquet} \\
\code{main_table_lpact_uncertainty_by_dataset_level.csv} & 9 & \code{controls/main_table_lpact_uncertainty_by_dataset_level.parquet} \\
\code{si_table_causal_uncertainty_by_mechanism.csv} & 18 & \code{controls/si_table_causal_uncertainty_by_mechanism.parquet} \\
\code{main_table_gate_threshold_sensitivity.csv} & 28 & \code{controls/main_table_gate_threshold_sensitivity.parquet} \\
\code{main_table_control_family_ablation.csv} & 13 & \code{controls/main_table_control_family_ablation.parquet} \\
\code{main_table_brain_as_model_oracle_gate.csv} & 2 & \code{controls/main_table_brain_as_model_oracle_gate.parquet} \\
\code{main_table_leave_one_dataset_out_summary.csv} & 3 & \code{controls/main_table_leave_one_dataset_out_summary.parquet} \\
\code{main_table_leave_one_subject_run_out_summary.csv} & 7 & \code{controls/main_table_leave_one_subject_run_out_summary.parquet} \\
\code{main_table_model_inventory.csv} & 9 & \code{controls/main_table_model_inventory.parquet} \\
\bottomrule
\end{longtable}

\section{Limitations Recorded by the SI}

The control-explained interpretation is bounded by the analyzed derived artifacts, preprocessing, and model coverage. Relational profile deltas are concentrated in MEG-MASC because that is where the locked analysis package provides the required matchable model-feature and brain-unit grid. Brain Treebank and Podcast ECoG contribute predictive and mechanism-stripping closure rows, but the integrated gates do not pass. Ceiling estimates are valid only where repeated-measure structure permits the configured reliability calculation.

These limitations define the scope of the main claim: prediction is not alignment in the analyzed derived artifact set. The result does not rule out future models, prospective neural data, richer annotations, or stronger repeated-measure designs. It shows that the present source-audited evidence, with severe controls and reliability bounds, does not support positive structural or mechanism-specific brain-language-model alignment.

\section{Scope and Limitation FAQ}

\begingroup
\small
\setlength{\tabcolsep}{3pt}
\begin{longtable}{L{4.1cm} L{10.0cm}}
\caption{Scope and limitation FAQ. These answers summarize how the manuscript should be read and what the available evidence does and does not support.}\\
\toprule
Question & Boundary answer \\
\midrule
\endfirsthead
\toprule
Question & Boundary answer \\
\midrule
\endhead
How conservative is L-PACT? & L-PACT is conservative for structural and mechanism-specific claims by design. Less stringent positive rows are retained in conventional-contrast and row-level tables, but they are not promoted to alignment when severe controls, relational profiles, mechanism stripping, or reliability bounds do not pass. \\
Why are all real model rows nonpassing? & Assay-sensitivity controls distinguish model-specific nonpasses from failures of the assay itself. Brain-brain, brain-as-model, low-level neural, acoustic-envelope, and synthetic implanted-signal controls show positive outcomes where stable brain-derived, low-level, acoustic, or known implanted structure is present. \\
What evidence would count as positive? & A positive integrated L-PACT candidate would need the same row or claim unit to pass predictive, relational, mechanism-stripping, reliability-bounded, severe-control, and replication gates, with the relevant source chain present. \\
Could one severe control be too strong? & Control-family ablation shows that removing circular-shift, autocorrelation-matched random, or layer-label controls does not create an integrated positive row. Apparent counts increase mainly when the rule ignores the corresponding severe contrast. \\
Why exclude NaturalStories from primary neural evidence? & NaturalStories is retained as diagnostic or stimulus-LLM-only because the available derived artifact chain does not satisfy the configured neural-side source, control, and reliability requirements for primary L-PACT inference. \\
Why not conclude that language models never align with brains? & The result is bounded to analyzed derived artifacts, preprocessing, controls, and the tested model representations. It does not rule out future models, prospective neural data, richer annotations, or stronger repeated-measure designs. \\
Why use derived artifacts rather than reprocess all raw data? & The manuscript is a source-audited reanalysis of locked derived neural and model artifacts. This preserves the estimand, prevents post hoc model searching, and respects the boundary that raw neural data and stimulus media are not redistributed. \\
Why not add larger models now? & Adding new models would change the study from a bounded evidence-standard evaluation into an open-ended model sweep. The inventory records larger or missing registry models as coverage boundaries, and future prospective analyses can test them against the same gates. \\
\bottomrule
\end{longtable}
\endgroup

\section{Reproducibility Notes}

The final reproducibility record includes locked Parquet tables, CSV exports, schema JSON files, SHA256 checksums, provenance records, and reproducibility scripts. The manuscript table exports additionally include row-level CSV copies under \code{manuscript/tables/detailed/}. The claim-boundary tables report no unsupported claim warnings. Checksum and provenance checks establish integrity and traceability; they are not scientific gates and do not add positive alignment evidence.


\begin{thebibliography}{99}
\bibitem{huth2016} A. G. Huth, W. A. de Heer, T. L. Griffiths, F. E. Theunissen, J. L. Gallant, Natural speech reveals the semantic maps that tile human cerebral cortex. \textit{Nature} \textbf{532}, 453--458 (2016).
\bibitem{jain2018} S. Jain, A. G. Huth, Incorporating context into language encoding models for fMRI. \textit{Advances in Neural Information Processing Systems} \textbf{31} (2018).
\bibitem{schrimpf2021} M. Schrimpf et al., The neural architecture of language: Integrative modeling converges on predictive processing. \textit{Proceedings of the National Academy of Sciences} \textbf{118}, e2105646118 (2021).
\bibitem{caucheteux2022} C. Caucheteux, J.-R. King, Brains and algorithms partially converge in natural language processing. \textit{Communications Biology} \textbf{5}, 134 (2022).
\bibitem{goldstein2022} A. Goldstein et al., Shared computational principles for language processing in humans and deep language models. \textit{Nature Neuroscience} \textbf{25}, 369--380 (2022).
\bibitem{toneva2019} M. Toneva, L. Wehbe, Interpreting and improving natural-language processing in machines with natural language-processing in the brain. \textit{Advances in Neural Information Processing Systems} \textbf{32} (2019).
\bibitem{tuckute2024} G. Tuckute et al., Driving and suppressing the human language network using large language models. \textit{Nature Human Behaviour} \textbf{8}, 544--561 (2024).
\bibitem{pereira2018} F. Pereira et al., Toward a universal decoder of linguistic meaning from brain activation. \textit{Nature Communications} \textbf{9}, 963 (2018).
\bibitem{wehbe2014} L. Wehbe et al., Simultaneously uncovering the patterns of brain regions involved in different story reading subprocesses. \textit{PLOS ONE} \textbf{9}, e112575 (2014).
\bibitem{lerner2011} Y. Lerner, C. J. Honey, L. J. Silbert, U. Hasson, Topographic mapping of a hierarchy of temporal receptive windows using a narrated story. \textit{Journal of Neuroscience} \textbf{31}, 2906--2915 (2011).
\bibitem{fedorenko2011} E. Fedorenko, A. Behr, N. Kanwisher, Functional specificity for high-level linguistic processing in the human brain. \textit{Proceedings of the National Academy of Sciences} \textbf{108}, 16428--16433 (2011).
\bibitem{blank2016} I. Blank, Z. Balewski, K. Mahowald, E. Fedorenko, Syntactic processing is distributed across the language system. \textit{NeuroImage} \textbf{127}, 307--323 (2016).
\bibitem{brennan2016} J. R. Brennan, E. P. Stabler, S. E. Van Wagenen, W.-M. Luh, J. T. Hale, Abstract linguistic structure correlates with temporal activity during naturalistic comprehension. \textit{Brain and Language} \textbf{157--158}, 81--94 (2016).
\bibitem{ding2016} N. Ding, L. Melloni, H. Zhang, X. Tian, D. Poeppel, Cortical tracking of hierarchical linguistic structures in connected speech. \textit{Nature Neuroscience} \textbf{19}, 158--164 (2016).
\bibitem{brodbeck2018} C. Brodbeck, A. Presacco, J. Z. Simon, Rapid transformation from auditory to linguistic representations of continuous speech. \textit{Current Biology} \textbf{28}, 3976--3983.e5 (2018).
\bibitem{broderick2018} M. P. Broderick, A. J. Anderson, G. M. Di Liberto, M. J. Crosse, E. C. Lalor, Electrophysiological correlates of semantic dissimilarity reflect the comprehension of natural, narrative speech. \textit{Current Biology} \textbf{28}, 803--809.e3 (2018).
\bibitem{brainscore} M. Schrimpf et al., Brain-Score: a benchmark for neural predictivity of artificial visual systems. \textit{bioRxiv}, 407007 (2018).
\bibitem{yamins2014} D. L. K. Yamins et al., Performance-optimized hierarchical models predict neural responses in higher visual cortex. \textit{Proceedings of the National Academy of Sciences} \textbf{111}, 8619--8624 (2014).
\bibitem{mikolov2013} T. Mikolov, I. Sutskever, K. Chen, G. Corrado, J. Dean, Distributed representations of words and phrases and their compositionality. \textit{Advances in Neural Information Processing Systems} \textbf{26} (2013).
\bibitem{pennington2014} J. Pennington, R. Socher, C. D. Manning, GloVe: global vectors for word representation. \textit{Proceedings of the 2014 Conference on Empirical Methods in Natural Language Processing}, 1532--1543 (2014).
\bibitem{vaswani2017} A. Vaswani et al., Attention is all you need. \textit{Advances in Neural Information Processing Systems} \textbf{30} (2017).
\bibitem{devlin2019} J. Devlin, M.-W. Chang, K. Lee, K. Toutanova, BERT: pre-training of deep bidirectional transformers for language understanding. \textit{Proceedings of NAACL-HLT}, 4171--4186 (2019).
\bibitem{radford2019} A. Radford et al., Language models are unsupervised multitask learners. \textit{OpenAI Technical Report} (2019).
\bibitem{brown2020} T. B. Brown et al., Language models are few-shot learners. \textit{Advances in Neural Information Processing Systems} \textbf{33}, 1877--1901 (2020).
\bibitem{kaplan2020} J. Kaplan et al., Scaling laws for neural language models. \textit{arXiv}:2001.08361 (2020).
\bibitem{hoffmann2022} J. Hoffmann et al., Training compute-optimal large language models. \textit{arXiv}:2203.15556 (2022).
\bibitem{biderman2023} S. Biderman et al., Pythia: a suite for analyzing large language models across training and scaling. \textit{Proceedings of the 40th International Conference on Machine Learning}, 2397--2430 (2023).
\bibitem{qwen25} A. Yang et al., Qwen2.5 Technical Report. \textit{arXiv}:2412.15115 (2024).
\bibitem{qwen3} A. Yang et al., Qwen3 Technical Report. \textit{arXiv}:2505.09388 (2025).
\bibitem{naselaris2011} T. Naselaris, K. N. Kay, S. Nishimoto, J. L. Gallant, Encoding and decoding in fMRI. \textit{NeuroImage} \textbf{56}, 400--410 (2011).
\bibitem{hastie2009} T. Hastie, R. Tibshirani, J. Friedman, \textit{The Elements of Statistical Learning}, 2nd ed. (Springer, 2009).
\bibitem{stone1974} M. Stone, Cross-validatory choice and assessment of statistical predictions. \textit{Journal of the Royal Statistical Society: Series B} \textbf{36}, 111--133 (1974).
\bibitem{kriegeskorte2008} N. Kriegeskorte, M. Mur, P. Bandettini, Representational similarity analysis: connecting the branches of systems neuroscience. \textit{Frontiers in Systems Neuroscience} \textbf{2}, 4 (2008).
\bibitem{kornblith2019} S. Kornblith, M. Norouzi, H. Lee, G. Hinton, Similarity of neural network representations revisited. \textit{Proceedings of the 36th International Conference on Machine Learning}, 3519--3529 (2019).
\bibitem{apa} L. Hoefling, M. Tangemann, L. Piefke, S. Keller, M. Bethge, K. Franke, Only Brains Align with Brains: cross-region alignment patterns expose limits of normative models. \textit{International Conference on Learning Representations (ICLR)}, poster. OpenReview:cMGJcHHI7d (2026).
\bibitem{neuroaituring} J. Feather, M. Khosla, N. A. R. Murty, A. Nayebi, Brain-model evaluations need the NeuroAI Turing Test. \textit{arXiv}:2502.16238 (2025).
\bibitem{cadieu2014} C. F. Cadieu et al., Deep neural networks rival the representation of primate IT cortex for core visual object recognition. \textit{PLOS Computational Biology} \textbf{10}, e1003963 (2014).
\bibitem{khaligh2014} S.-M. Khaligh-Razavi, N. Kriegeskorte, Deep supervised, but not unsupervised, models may explain IT cortical representation. \textit{PLOS Computational Biology} \textbf{10}, e1003915 (2014).
\bibitem{yamins2016} D. L. K. Yamins, J. J. DiCarlo, Using goal-driven deep learning models to understand sensory cortex. \textit{Nature Neuroscience} \textbf{19}, 356--365 (2016).
\bibitem{kell2018} A. J. E. Kell, D. L. K. Yamins, E. N. Shook, S. V. Norman-Haignere, J. H. McDermott, A task-optimized neural network replicates human auditory behavior, predicts brain responses, and reveals a cortical processing hierarchy. \textit{Neuron} \textbf{98}, 630--644.e16 (2018).
\bibitem{richards2019} B. A. Richards et al., A deep learning framework for neuroscience. \textit{Nature Neuroscience} \textbf{22}, 1761--1770 (2019).
\bibitem{algonauts} R. M. Cichy et al., The Algonauts Project 2021 Challenge: How the human brain makes sense of a world in motion. \textit{arXiv}:2104.13714 (2021).
\bibitem{nastase2021} S. A. Nastase et al., The ``Narratives'' fMRI dataset for evaluating models of naturalistic language comprehension. \textit{Scientific Data} \textbf{8}, 250 (2021).
\bibitem{gorgolewski2016} K. J. Gorgolewski et al., The brain imaging data structure, a format for organizing and describing outputs of neuroimaging experiments. \textit{Scientific Data} \textbf{3}, 160044 (2016).
\bibitem{markiewicz2021} C. J. Markiewicz et al., The OpenNeuro resource for sharing of neuroscience data. \textit{eLife} \textbf{10}, e71774 (2021).
\bibitem{gramfort2013} A. Gramfort et al., MEG and EEG data analysis with MNE-Python. \textit{Frontiers in Neuroscience} \textbf{7}, 267 (2013).
\bibitem{poldrack2017} R. A. Poldrack et al., Scanning the horizon: towards transparent and reproducible neuroimaging research. \textit{Nature Reviews Neuroscience} \textbf{18}, 115--126 (2017).
\bibitem{hoerl1970} A. E. Hoerl, R. W. Kennard, Ridge regression: biased estimation for nonorthogonal problems. \textit{Technometrics} \textbf{12}, 55--67 (1970).
\bibitem{pedregosa2011} F. Pedregosa et al., Scikit-learn: machine learning in Python. \textit{Journal of Machine Learning Research} \textbf{12}, 2825--2830 (2011).
\bibitem{varma2006} S. Varma, R. Simon, Bias in error estimation when using cross-validation for model selection. \textit{BMC Bioinformatics} \textbf{7}, 91 (2006).
\bibitem{varoquaux2017} G. Varoquaux et al., Assessing and tuning brain decoders: cross-validation, caveats, and guidelines. \textit{NeuroImage} \textbf{145}, 166--179 (2017).
\bibitem{yarkoni2017} T. Yarkoni, J. Westfall, Choosing prediction over explanation in psychology: lessons from machine learning. \textit{Perspectives on Psychological Science} \textbf{12}, 1100--1122 (2017).
\bibitem{mantel1967} N. Mantel, The detection of disease clustering and a generalized regression approach. \textit{Cancer Research} \textbf{27}, 209--220 (1967).
\bibitem{haxby2001} J. V. Haxby et al., Distributed and overlapping representations of faces and objects in ventral temporal cortex. \textit{Science} \textbf{293}, 2425--2430 (2001).
\bibitem{cortes2012} C. Cortes, M. Mohri, A. Rostamizadeh, Algorithms for learning kernels based on centered alignment. \textit{Journal of Machine Learning Research} \textbf{13}, 795--828 (2012).
\bibitem{gretton2005} A. Gretton, O. Bousquet, A. Smola, B. Schoelkopf, Measuring statistical dependence with Hilbert-Schmidt norms. \textit{Algorithmic Learning Theory}, 63--77 (2005).
\bibitem{kriegeskorte2009} N. Kriegeskorte, W. K. Simmons, P. S. F. Bellgowan, C. I. Baker, Circular analysis in systems neuroscience: the dangers of double dipping. \textit{Nature Neuroscience} \textbf{12}, 535--540 (2009).
\bibitem{efron1994} B. Efron, R. J. Tibshirani, \textit{An Introduction to the Bootstrap} (Chapman and Hall/CRC, 1994).
\bibitem{nichols2002} T. E. Nichols, A. P. Holmes, Nonparametric permutation tests for functional neuroimaging: a primer with examples. \textit{Human Brain Mapping} \textbf{15}, 1--25 (2002).
\bibitem{benjamini1995} Y. Benjamini, Y. Hochberg, Controlling the false discovery rate: a practical and powerful approach to multiple testing. \textit{Journal of the Royal Statistical Society: Series B} \textbf{57}, 289--300 (1995).
\end{thebibliography}
\end{document}